\documentclass[final]{jpp}
\usepackage{graphicx}
\usepackage[utf8]{inputenc}
\usepackage[T1]{fontenc}
\usepackage{caption}
\usepackage{amsmath, amssymb}
\usepackage{bm}
\usepackage{upgreek}
\usepackage{xcolor}
\usepackage{hyperref}
\usepackage{amsmath}
\usepackage{subfigure}
\newcommand{\answer}[1]{\textcolor{black}{#1}}

\shorttitle{Plasma pressure response to non-inductive current drive}
\shortauthor{A. Krupka and M.-C. Firpo}

\title{Plasma pressure response to non-inductive current drive in axisymmetric visco-resistive MHD steady-states}

\author{Anna Krupka\aff{1} \corresp{\email{anna.krupka@lpp.polytechnique.fr}}, 
  Marie-Christine Firpo\aff{1} \corresp{\email{marie-christine.firpo@lpp.polytechnique.fr}}}

\affiliation{\aff{1}Laboratoire de Physique des Plasmas (LPP), CNRS, Sorbonne Universit\'{e}, Ecole polytechnique, Institut Polytechnique de Paris, 91120 Palaiseau, France}

\begin{document}

\maketitle
\thispagestyle{plain}

\begin{abstract}
We investigate self-consistent, steady-state axisymmetric solutions of incompressible tokamak plasma using a visco-resistive magnetohydrodynamic model. A key contribution of this work is the formulation of Poisson's equation that governs the pressure profile. Our analysis reveals that the current modeling fails to produce realistic pressure levels. To overcome this limitation, we introduce additional non-inductive current drives, akin to those generated by neutral beam injection or radio frequency heating, modeled as modifications to the toroidal current. Numerical simulations validate our enhanced model, showing significant improvements in pressure profile characteristics. In the cases examined, the effect of these current drives on the velocity profiles is moderate, except when the non-inductive current drives induce reversals in the total toroidal current density, leading to non-nested flux surfaces with internal separatrices.
\end{abstract}


\section{\label{Introduction}Introduction}

The calculation of plasma equilibria is a critical aspect of both the design and operation of magnetic confinement devices used in nuclear fusion research. In particular, a thorough understanding of tokamak plasma physics within the axisymmetric framework is crucial, as it serves as the foundation from which three-dimensional perturbations inevitably arise. From a numerical point of view, among the most realistic models are those that (gyro-)kinetically describe each plasma species, coupled with Maxwell's equations and incorporating the external driving forces present in tokamak devices. These are known to be computationally intensive and complex. Furthermore, implementing boundary conditions in such kinetic models presents significant challenges \citep{Ball2020}. In real experiments, the Grad-Shafranov equation serves usually to reconstruct equilibria. This comes from the steady-state zero-flow and ideal Navier-Stokes equation. In the present study, our aim is to reintroduce self-consistency in this reconstruction without entering the difficulties and details of a gyrokinetic approach. This amounts to using a magnetohydrodynamic (MHD) approach. We then consider a steady-state axisymmetric (2D) visco-resistive MHD closed set of equations compatible with tokamak conditions.

In this frame, a key challenge in developing a minimal, yet meaningful, model lies in correctly representing the physical drives at work within the device. The \emph{first natural drive} is the curl-free magnetic field created by the external coils. A \emph{second drive} must be implemented to induce the winding of the magnetic field lines around the magnetic axis by creating a poloidal component of the magnetic field. \answer{In various previous works \citep{kamp2003toroidal, KAMP_MONTGOMERY_2004, oueslati2019numerical}, this second drive has been an external toroidal field produced by the time variation of the poloidal magnetic flux, assumed to be a non-zero constant, which makes the system time independent.} Under the constraint of axisymmetry, the stationary plasma states can then be determined by solving a self-consistent system consisting of the steady-state Navier–Stokes equation, the steady-state Maxwell equations with the two external drives, and Ohm’s law, which provides closure to the system. This system can be solved, for example, using the finite element method, once the plasma domain $\Omega$ and the boundary conditions for the fields have been specified. It has been studied \citep{kamp2003toroidal, KAMP_MONTGOMERY_2004, oueslati2019numerical} primarily with the aim of estimating plasma flow velocities in steady-state regimes. It has recently been shown \citep{KRUPKA_scaling} that the dependence of the visco-resistive system on the two external drives can be reformulated as a dependence on a single control parameter. This amounts to the ratio of the electric current driven by Ohm's law in response to an applied toroidal loop voltage over that needed for generating the external toroidal magnetic field. A second relevant control parameter relative to this visco-resistive framework is the Hartmann number defined as $H=(\eta \nu)^{-1/2}$, where $\eta$ and $\nu$ are respectively the dimensionless plasma resistivity and viscosity. 

\answer{In the present study, another path to produce the \emph{second drive} will be implemented in the form of a non-inductive current drive. This is valuable for investigating advanced operational regimes that are relevant for the truly steady-state operation of tokamaks \citep{Kikuchi2010}.  Long-duration, steady-state-like operation has already been demonstrated in several existing tokamaks \citep{Houtte_2004, Ferron_2013, NBI_EAST_2023, Ko_2024}, providing important benchmarks for future devices designed for fully steady-state scenarios, including DEMO \citep{TRAN2022}, CFETR \citep{Wan_2017} or JT-60SA \citep{Garzotti_2018}.}

In this self-consistent framework, \answer{irrespective of the inductive or non-inductive nature of the toroidal current drive},  the plasma pressure field has been largely disregarded since it can be eliminated by taking the curl of the steady-state Navier–Stokes equation. This reflects the well-established notion, familiar in the study of the Navier-Stokes equation applied to neutral fluids, that pressure behaves as a passive rather than an active variable. In the present work, however, we focus on the evaluation of the pressure field, an aspect that has been overlooked within this approach so far. The derivation provided here is fully self-consistent, marking a clear departure from the conventional treatment of pressure in tokamak plasma modeling. Indeed, conventional approaches, such as real-time equilibrium reconstruction codes using the Grad-Shafranov equation, or its extended versions incorporating some plasma flows, also known as the Grad–Shafranov–Bernoulli system of equations \citep{Guazzotto2004,Li2021,DelPrete2021,Kaltsas2022,Torija_Daza_2024}, treat the scalar pressure field as a free function. This function, along with the diamagnetic function (and possibly functions associated to plasma flows), is optimized to minimize the $\chi^{2}$ from the measured data. In other typical models for tokamak plasmas, the pressure may be evaluated using an equation of state, which amounts to a thermodynamic closure.

This paper is organized as follows. In Section \ref{Original system}, we introduce the aforementioned steady-state axisymmetric model for describing tokamak plasmas within an incompressible visco-resistive MHD framework having as time-independent external drives a curl-free toroidal magnetic field and a curl-free toroidal electric field \citep{kamp2003toroidal, KAMP_MONTGOMERY_2004, oueslati2019numerical,KRUPKA_scaling}. Specifically, it is predicted that this system yields a zero pressure gradient in the ideal and motionless limit. This points to the necessity of incorporating an additional non-inductive current drive in a steady-state machine to effectively control and increase the pressure. This is addressed in Section \ref{Implementation} where we implement some current drive to model the heating methods used in real tokamaks verified through pressure profiles in Section \ref{Pressure}. Numerical simulations with toroidal current drives are presented in Section \ref{Numerics}, using the finite element method through the FreeFem$++$ platform for solving partial differential equations~\citep{hecht2012new}. In conclusion, Section \ref{Conclusion} summarizes the study's findings.

\section{\label{Original system} The necessity of non-inductive current drive: A theoretical approach}

\subsection{\label{Sys_eq} Axisymmetric steady-state visco-resistive MHD: self-consistent system of equations}

The framework used in this study is magnetohydrodynamics. In more precise terms, building on the research initiated by Montgomery and his collaborators \citep{Kamp1998}, we assume that the axisymmetric steady-states of the plasma are governed by the incompressible visco-resistive MHD. This is consistent with the customary reconstruction of 2D equilibria using the Grad-Shafranov equation, except that we do not assume the velocity field to be zero, and we have a self-consistent model, as we do not have free functions. Then, to describe a tokamak plasma, an essential aspect is to model the external drives involved in the system. One inherent drive in this magnetic confinement fusion device is the external magnetic field. Additionally, the need to wind the magnetic field lines and create a macroscopic poloidal component of the magnetic field requires a second forcing mechanism. Following previous references \citep{kamp2003toroidal,KAMP_MONTGOMERY_2004,oueslati2019numerical, KRUPKA_scaling, oueslati2020breaking}, we assume that the poloidal magnetic field component is generated by a toroidal electric field, which drives a toroidal current density.

Denoting by $B_{0}$ the value of the external magnetic field on the magnetic axis, by $\mu_{0}$ the vacuum permeability and by $\rho_{m0}$ the plasma mass density assumed to be constant, the Alfv\'{e}n velocity is $v_{A0}=B_{0}/(\mu_{0}\rho_{m0})^{1/2}$. In the remainder of this article, we shall work with dimensionless variables. Specifically, velocities are normalized with respect to the Alfvén speed, $v_{A0}$, as is the field $\boldsymbol{B}/(\mu_{0}\rho_{m0})^{1/2}$. Moreover, the space variables are also dimensionless. From the set of cylindrical polar coordinates $(r,\varphi,z)$ and denoting by $R_{0}$ the tokamak major radius, we define $R=r/R_{0}$ and $Z=z/R_{0}$. The dimensionless resistivity, $\eta$, is the resistivity divided by $\mu_{0} R_{0} v_{A0}$ and the dimensionless viscosity, $\nu$, is the viscosity divided by $R_{0} v_{A0}$. The computation of visco-resistive axisymmetric steady states involves then solving the steady-state incompressible Navier–Stokes equation (\ref{N_S})-(\ref{Inc}) along with the solenoidal condition (\ref{Max1}), Amp\`{e}re law (\ref{Max3}) and Ohm's law (\ref{Ohm}) on a tokamak poloidal plasma cross-section $\Omega$. The equations are
\begin{align}
    \label{N_S}
&(\boldsymbol{v}\cdot\nabla)\boldsymbol{v}=\boldsymbol{J}\times\boldsymbol{B}-\nabla p+\nu\nabla^2\boldsymbol{v}, \\
    \label{Inc}
&\nabla\cdot\boldsymbol{v}=0,\\
\label{Max1}
&\nabla\cdot\boldsymbol{B}=0,\\
\label{Max3}
&\nabla\times\boldsymbol{B}=\boldsymbol{J},\\
\label{Ohm}
&\boldsymbol{E}+\boldsymbol{v}\times\boldsymbol{B}=\eta\boldsymbol{J}.
\end{align}
With respect to the drives, the externally applied (vacuum) toroidal magnetic field is
\begin{equation}
\boldsymbol{B}_{\mathrm{ext}}=\frac{1}{R}\boldsymbol{i}_{\varphi},\label{external_fields_B}
\end{equation}
with $\boldsymbol{i}_{\varphi}$ a unit vector in the toroidal (azimuthal) direction. \answer{As for the electric field, its general expression is
\begin{equation}
\mathbf{E}=-\mathbf{\nabla }\Phi +V\mathbf{\nabla }\left( \frac{\varphi }{%
2\pi }\right),
\end{equation}
where $-\mathbf{\nabla }\Phi $ is a purely electrostatic contribution and where $V$ is the loop voltage. This second term serves to inductively generate the toroidal current, it is, therefore, non-steady-state in nature and amounts to an external drive
\begin{equation}
\mathbf{E}_{\mathrm{ext}}=\frac{E_{0}}{R}\mathbf{i}_{\varphi } \label{E_ext}
\end{equation}
with $E_{0}=(2\pi )^{-1}\partial \psi _{pol}/\partial t$. In the present modeling, we consider the time variation of the poloidal magnetic flux $\psi_{\mathrm{pol}}$, and thus $E_{0}$, as constant and nonzero. This field satisfies $\mathbf{\nabla \times E}_{\mathrm{ext}}=\mathbf{0}$ pointwise for $R>0$, but its
global structure exhibits a non-vanishing toroidal circulation, consistent with a finite $V=\partial \psi_{\mathrm{pol}}/\partial t$. Consequently, although $\mathbf{E}_{\mathrm{ext}}$ is locally curl-free in the classical sense, it does not derive from a single-valued scalar potential, and Faraday's law must be
interpreted in the weak (distributional) sense. Thus, the present frame is strictly speaking not steady state as it is required that $\partial \psi_{\mathrm{pol}}/\partial t$ be nonzero, but it is time independent.  Nevertheless, once the current drive is implemented (see Sec. \ref{Pressure_field_behaviour}), we can
put $E_{0}=0$ in the model equations and just have $\mathbf{E}=-\mathbf{\nabla }\Phi.$} Let us note, indeed, that several tokamak experiments have demonstrated scenarios where the externally applied electric field is curl-free  while maintaining a \answer{zero} toroidal component (\ref{E_ext}).
This occurs in fully non-inductive discharges driven by mechanisms such as Lower Hybrid Current Drive (LHCD), Electron Cyclotron Current Drive (ECCD), and Neutral Beam Current Drive (NBCD), where the electric field is electrostatic or wave-driven, not induced by transformer action (See e.g. \cite{Sauter_2000,Litaudon_2002}). The magnetic and electric fields in Eqs. (\ref{N_S})-(\ref{Ohm}) are the sum of these external contributions and of the self-consistent plasma fields. This system of equations needs to be solved in the plasma cross-section $\Omega$ with suitable boundary conditions. From a computational perspective, we solve the system of partial differential equations (PDE) that we are now presenting.

\subsection{Scalar PDE formulation}
One can eliminate the unknown pressure term by taking the curl of Eq. (\ref{N_S}). This signifies that the pressure is not an active but a passive variable. Moreover, the single-fluid, mass-averaged plasma velocity $\boldsymbol{v}$, vorticity $\boldsymbol{\omega}\equiv \nabla \times \boldsymbol{v}$, magnetic $\boldsymbol{B}$ and current density $\boldsymbol{J}$ vector fields are divergence-free, they admit then the following representations
\begin{align}
\boldsymbol{v} &= \frac{1}{R}\nabla\chi\times\boldsymbol{i}_{\varphi}+v_{\varphi }\boldsymbol{i}_{\varphi} ,\\
\boldsymbol{\omega}&=\frac{1}{R}\nabla \left( Rv_{\varphi }\right) \times \boldsymbol{i}_{\varphi}-\frac{1}{R}\left( \triangle^{\ast }\chi\right) \boldsymbol{i}_{\varphi},\label{vorticity}\\
\boldsymbol{B} &= \frac{1}{R} \nabla \psi \times \boldsymbol{i}_{\varphi} + B_{\varphi} \boldsymbol{i}_{\varphi},\\
\boldsymbol{J} &=\frac{1}{R}\nabla\left( RB_{\varphi}\right) \times \boldsymbol{i}_{\varphi}-\frac{1}{R}\left(
        \triangle^{\ast }\psi\right) \boldsymbol{i}_{\varphi} \label{current},
\end{align}
where $\chi$ is the velocity stream function, $\psi$ is the magnetic flux function, $B_{\varphi}$ is the toroidal component of the magnetic field vector and $v_{\varphi }$ is the toroidal component of the velocity field vector. The above system of equations (\ref{N_S})-(\ref{Ohm}) with the external drives  (\ref{external_fields_B})-(\ref{E_ext}) can be expressed~\cite{kamp2003toroidal, Kamp1998, KAMP_MONTGOMERY_2004,KRUPKA_scaling, roverchSteadystateFlowsViscoresistive2021} as the following set of five scalar elliptic PDE
\begin{align}
    \label{eq:set_equation1}
\triangle ^{\ast }\chi &=-R\omega_{\varphi}, \\ \nonumber
    \label{eq:set_equation2}
  \nu\triangle ^{\ast }(R\omega_{\varphi})&=\frac{1}{R^{2}}\frac{\partial }{\partial
Z}((RB_{\varphi})^{2}-(Rv_{\varphi})^{2})+\frac{1}{R}\left\{ Rj_{\varphi},\psi\right\}\\
&+\frac{1}{R}\left\{ \chi,R\omega_{\varphi}\right\}+
 \frac{2\omega_{\varphi}}{R}\frac{\partial \chi}{%
\partial Z}-\frac{2j_{\varphi}}{R}\frac{\partial \psi}{\partial Z},\\ \nonumber
\label{eq:set_equation3}
\eta\triangle ^{\ast }(RB_{\varphi})&=\frac{1}{R}\left\{ \chi,RB_{\varphi}\right\} +%
\frac{1}{R}\left\{ Rv_{\varphi},\psi\right\}+ \\&+\frac{2RB_{\varphi}}{R^{2}}\frac{\partial
\chi}{\partial Z}-\frac{2v_{\varphi}}{R}\frac{\partial \psi}{\partial Z},\\
\label{eq:set_equation4}
\nu\triangle ^{\ast }(Rv_{\varphi}) &=\frac{1}{R}\left\{ \chi,Rv_{\varphi}\right\} +%
\frac{1}{R}\left\{ RB_{\varphi},\psi\right\}, \\
\label{eq:set_equation5}
\triangle ^{\ast }\psi &=-Rj_{\varphi},
\end{align}
with the toroidal projection of Ohm's law giving the constraint
\begin{equation} \label{eq:set_equation6}
\eta Rj_{\varphi}=E_{0}+\frac{1}{R}\left\{ \psi,\chi\right\}.
\end{equation}
Here, the Poisson bracket $\left\{u,v\right\}$ for any spatial functions $u$ and $v$ is defined as
\begin{equation}
\left\{u,v\right\} \equiv \frac{\partial u}{\partial R}\frac{\partial v}{\partial Z}-\frac{\partial v}{\partial R}\frac{\partial u}{\partial Z},
\end{equation}
and the operator $\triangle ^{\ast }$ is defined by
\begin{equation}
\triangle ^{\ast } A \equiv \frac{\partial^{2}A}{\partial R^{2}} - \frac{1}{R} \frac{\partial A}{\partial R} +\frac{\partial^{2}A}{\partial Z^{2}}.
\end{equation}

A relevant dimensionless control parameter has been identified by ~\cite{Montgomery1993,Cappello,KRUPKA_scaling} as the Hartmann number, $H = (\eta \nu)^{-1/2}$. In fusion-relevant conditions, this is expected to be a large parameter, ranging from $10^6$ to $10^8$. Simulations for realistic parameter values already exist for this system under various boundary conditions, typically using the JET geometry. The elliptic system (\ref{eq:set_equation1})--(\ref{eq:set_equation5}) requires five boundary conditions. The four conditions associated with the divergence-free properties of the magnetic field ($\boldsymbol{B}$), current density ($\boldsymbol{J}$), velocity ($\boldsymbol{V}$), and vorticity ($\boldsymbol{\omega}$) vector fields are determined by ensuring the continuity of their normal components across the plasma boundary. The following boundary conditions are selected in the numerical simulations: $\chi = \psi = 0$ and $B_{\varphi} =1/R$ on $\partial \Omega$. We enforce Neumann boundary conditions on both the toroidal velocity $v_{\varphi}$ and toroidal vorticity $\omega_{\varphi}$ through $\partial_n v_{\varphi} = \partial_n \omega_{\varphi} = 0$ on $\partial \Omega$. We used the open-source PDE solver FreeFem++, employing the finite element method \cite{hecht2012new} to solve the above steady-state axisymmetric system of equations in a weak form on the plasma cross-section domain $\Omega$ with the specified boundary conditions. For our calculations, we set a tolerance parameter $\epsilon = 10^{-10}$, allowing the Newton-Raphson scheme to converge in typically 4--5 iterations.
\subsection{\label{Pres_grad}Examination of the pressure field}

Let us now examine the pressure profile in the visco-resistive model (\ref{N_S})-(\ref{Ohm}) with the external drives  (\ref{external_fields_B})-(\ref{E_ext}). Let us assume for now that the steady-state plasma speed is negligible. Then, in the ideal limit, $\eta \to 0$ and $\nu \to 0$, the steady-state Navier-Stokes equation (\ref{N_S}) takes the form
\begin{equation}
\mathbf{\nabla} p =\boldsymbol{J}\times \boldsymbol{B}. \label{EQGS}
\end{equation}%
Restricting to axisymmetric solutions, the projection of this equation on $R$ and $Z$ gives, respectively,
\begin{eqnarray}
\frac{\partial p}{\partial R}=R^{-1} \left( -B_{\varphi} \frac{\partial (RB_{\varphi})}{\partial R} + j_{\varphi} \frac{\partial \psi}{\partial R}\right),\label{gradp_r}\\
\frac{\partial p}{\partial Z}=R^{-1} \left( -B_{\varphi} \frac{\partial (RB_{\varphi})}{\partial Z} + j_{\varphi} \frac{\partial \psi}{\partial Z}\right).\label{gradp_z}
\end{eqnarray}
In the toroidal direction, we get
\begin{equation}
0=R^{-2} \{\psi,RB_{\varphi}\} \label{invo}
\end{equation}
which amounts to the well-known property of Grad-Shafranov's theory that the diamagnetic function, $R B_{\varphi}$, is a function of the magnetic flux $\psi$ only. Moreover, writing that $\boldsymbol{J}\times \boldsymbol{B}$ is curl-free, which follows from the force balance equation (\ref{EQGS}) and projecting this on the toroidal direction yields
\begin{equation}
-2 RB_{\varphi}\frac{\partial B_{\varphi}}{\partial Z} + \{\psi,Rj_{\varphi}\} +2 Rj_{\varphi} \frac{\partial \psi}{\partial Z}=0. \label{curljb}
\end{equation}
Then, combining Eqs. (\ref{gradp_z}) and (\ref{curljb}), the pressure gradient along the $z$-axis with a zero-flow hypothesis is given by
\begin{equation}
\frac{\partial p}{\partial Z}=\frac{1}{2 R} \left\{Rj_{\varphi},\psi\right\}.
\label{Pressure_gradient}
\end{equation}%
Yet, assuming no plasma flow, the toroidal projection of Ohm's law in Eq. (\ref{eq:set_equation6}) states that $Rj_{\varphi}$ is a constant, with $Rj_{\varphi} = E_{0}/\eta$. Eq. (\ref{Pressure_gradient}) indicates then that the pressure field does not depend on $Z$. However, from the set of equations (\ref{gradp_r})--(\ref{gradp_z})--(\ref{invo}), we can deduce that the pressure is a function of the magnetic flux $\psi$ such that $\{p, \psi\} = 0$. Thus, we have $\partial_{Z} p = p'(\psi) \partial_{Z} \psi = 0$. This implies that the pressure profile is constant. This aligns with the results obtained by \cite{Braams1991}, which indicate that in equilibrium configurations, the current density must be proportional to $1/R$ when the pressure gradient is zero.

\subsection{\label{Implementation}Implications and implementation of non-inductive current drives}

Section~\ref{Pres_grad} demonstrates that the sole inclusion of Ohm's law to close the system imposes significant limitations on the model. Specifically, the effective pressure in the model arises only because the toroidal geometry and viscous dissipation prevent the steady-state velocity field from being identically zero. This allows the pressure profile to remain non-zero, as $R j_{\varphi}$ is not exactly constant. However, the model lacks a robust mechanism to provide sufficient heating to achieve fusion conditions. Therefore, incorporating alternative heating methods is essential to reach higher pressure. We will also see that this tends to induce higher plasma rotation velocities in specific drive configurations.

In our previous analysis, we focused on the behaviour of the system for a specific ratio of $E_{0}/\eta$, which was the only explicit drive in the dimensionless system of equations \cite{KRUPKA_scaling}. Now, we aim to introduce an additional drive which will manifest as an extra term in the toroidal component of Ohm's law in Eq. (\ref{Ohm}) with
\begin{equation}
\frac{E_{0}}{R}\mathbf{+}\left( \boldsymbol{v}\times\boldsymbol{B}\right)
\cdot \boldsymbol{i}_{\varphi} + j_D= \eta\boldsymbol{J}\cdot \boldsymbol{i}_{\varphi},
\end{equation}
where $j_D$ represents a current drive. Eq. (\ref{eq:set_equation6}) now becomes
\begin{equation} \label{drive}
\eta Rj_{\varphi}=E_{0}+\frac{1}{R}\left\{ \psi,\chi\right\}+ j_D R. 
\end{equation}
Our goal is to investigate the influence of the non-inductive current drive $j_D$ within our system. We will begin by evaluating its effect on the pressure profile. For this purpose, it is necessary to determine the pressure field, which we will now establish.

\section{\label{Pressure}Derivation of Poisson's equation for the pressure field}

Let us go back to the steady-state Navier-Stokes equation (\ref{N_S}) and rewrite it as
\begin{equation}
    \boldsymbol{\omega}\times \boldsymbol{v}=\boldsymbol{J}\times\boldsymbol{B}-\nabla p^{\ast }+\nu\nabla^2\boldsymbol{v}
\label{N_S2}
\end{equation}
with%
\begin{equation}
p^{\ast }=p+\frac{{v}^{2}}{2}. \label{def_pstar}
\end{equation}
Previously, we eliminated the pressure term by taking the curl and considering the toroidal part of the force balance equation. Now, to obtain the pressure of the system, we take the divergence of Eq. (\ref{N_S2})
\begin{equation}
    \nabla\cdot\nabla p^{\ast }=\nabla\cdot\left[-\boldsymbol{\omega}\times \boldsymbol{v} +\boldsymbol{J}\times \boldsymbol{B}+\nu\nabla^{2}\boldsymbol{v}\right].
\end{equation}
This takes the form of Poisson's equation for the pressure $p^{\ast }$ as the left-hand side yields the Laplacian of the pressure, $\triangle p^{\ast}$. Taking the divergence of the first term on the right-hand side gives
\begin{equation}
   - \nabla\cdot(\boldsymbol{\omega}\times\boldsymbol{v})=\boldsymbol{v}\cdot\nabla^{2}\boldsymbol{v}+\boldsymbol{\omega}^{2}.
\end{equation}
We can treat the $\boldsymbol{J}\times\boldsymbol{B}$ term similarly
\begin{equation}
    \nabla\cdot(\boldsymbol{J}\times\boldsymbol{B})=-\boldsymbol{B}\cdot\nabla^{2}\boldsymbol{B}-\boldsymbol{J}^{2}.
\end{equation}
Finally, the term $\nabla\cdot(\nu\nabla^{2}\boldsymbol{v})$ equals zero due to the incompressibility condition $ \nabla\cdot\boldsymbol{v}=0$. Therefore, the complete Poisson's equation for the pressure is
\begin{equation}
    \triangle p^{\ast }=\boldsymbol{v}\cdot\nabla^{2}\boldsymbol{v}+\boldsymbol{\omega}^{2}-\boldsymbol{B}\cdot\nabla^{2}\boldsymbol{B}-\boldsymbol{J}^{2}
    \label{Pois_Pressure}
\end{equation}
where $\triangle$ is defined as
\begin{equation}
    \triangle A \equiv \frac{\partial^{2} A}{\partial R^{2}}+\frac{1}{R}\frac{\partial A}{\partial R}+\frac{\partial^{2} A}{\partial Z^{2}}.
\end{equation}
Next, we will express the Poisson's equation (\ref{Pois_Pressure}) in terms of the functions $\chi, \ldots, j_{\varphi}$ defined over the domain $(R,Z) \in \Omega$. To do this, we will analyze each term separately. By utilizing the expression for the vorticity (\ref{vorticity}), the second term can be rewritten as
\begin{equation}
    \boldsymbol{\omega}^{2}=-\frac{\omega_{\varphi}}{R}\triangle^{\ast}\chi+\left(\frac{1}{R}\frac{\partial (Rv_{\varphi})}{\partial R}\right)^2+\left(\frac{\partial v_{\varphi}}{\partial Z}\right)^2.
\end{equation}
Similarly, for the square of the current density vector (\ref{current}), we get
\begin{equation}
    \boldsymbol{J}^{2}=-\frac{j_{\varphi}}{R}\triangle^{\ast}\psi+\left(\frac{1}{R}\frac{\partial (RB_{\varphi})}{\partial R}\right)^2+\left(\frac{\partial B_{\varphi}}{\partial Z}\right)^2.
\end{equation}
Finally, let us examine the term $\boldsymbol{B} \cdot \nabla^2 \boldsymbol{B}$. We can use the identity $\boldsymbol{B} \cdot \nabla^2 \boldsymbol{B} = -\boldsymbol{B} \cdot (\nabla \times \boldsymbol{J})$ with
\begin{equation}
    \boldsymbol{B}\cdot(\nabla\times\boldsymbol{J})=-\frac{B_{\varphi}}{R}\triangle^{\ast}(RB_{\varphi}) + \frac{1}{R^{2}}\frac{\partial \psi}{\partial R}\frac{\partial (Rj_{\varphi})}{\partial R} + \frac{1}{R}\frac{\partial \psi}{\partial Z}\frac{\partial j_{\varphi}}{\partial z}.
\end{equation}
Similarly, we have
\begin{equation}
    \boldsymbol{v}\cdot(\nabla\times\boldsymbol{\omega})=-\frac{v_{\varphi}}{R}\triangle^{\ast}(Rv_{\varphi}) + \frac{1}{R^{2}}\frac{\partial (R\omega_{\varphi})}{\partial R}\frac{\partial \chi}{\partial R} + \frac{1}{R}\frac{\partial \omega_{\varphi}}{\partial Z}\frac{\partial \chi}{\partial Z}.
\end{equation}
Incorporating all of these contributions into the right-hand side of Poisson's equation yields
\begin{align}
\triangle p^{\ast } &= \frac{v_{\varphi}}{R}\triangle^{\ast}(Rv_{\varphi}) - \frac{1}{R^{2}}\frac{\partial (R\omega_{\varphi})}{\partial R}\frac{\partial \chi}{\partial R} - \frac{1}{R}\frac{\partial \omega_{\varphi}}{\partial Z}\frac{\partial \chi}{\partial Z} \nonumber \\
&-\frac{\omega_{\varphi}}{R}\triangle^{\ast}\chi+\left(\frac{1}{R}\frac{\partial (Rv_{\varphi})}{\partial R}\right)^2+\left(\frac{\partial v_{\varphi}}{\partial Z}\right)^2\nonumber \\
& -\frac{B_{\varphi}}{R}\triangle^{\ast}(RB_{\varphi}) + \frac{1}{R^{2}}\frac{\partial \psi}{\partial R}\frac{\partial (Rj_{\varphi})}{\partial R} + \frac{1}{R}\frac{\partial \psi}{\partial Z}\frac{\partial j_{\varphi}}{\partial Z}\nonumber \\
& +\frac{j_{\varphi}}{R}\triangle^{\ast}\psi-\left(\frac{1}{R}\frac{\partial (RB_{\varphi})}{\partial R}\right)^2-\left(\frac{\partial B_{\varphi}}{\partial Z}\right)^2.
\label{Pressure_poisson}
\end{align}
This elliptic differential equation needs to be solved with a boundary condition, and allow the pressure profiles for the different drives to be computed. To the best of our knowledge, the derivation of Eq. (\ref{Pressure_poisson}) within the visco-resistive system is novel.

\section{\label{Numerics}Numerical results}

\subsection{\label{Pressure_field_behaviour}Pressure field behavior without and with non-inductive current drives}

Let us now solve Poisson's equation for pressure, assuming a zero pressure condition at the boundary $\partial \Omega$. It is important to note that, in the absence of an additional toroidal current drive and assuming no plasma flow, we previously inferred a zero pressure gradient, resulting in a constant zero pressure in the limits $\eta \to 0$ and $\nu \to 0$, as discussed in Section \ref{Pres_grad}. To return to dimensional pressure and compare the results with those obtained from the JET tokamak, we recall that $p^{\ast}$ is the dimensionless total pressure \cite{KRUPKA_scaling}, normalized as $p^{\ast} = \hat{p}^{\ast}/v_{A0}^2 \rho_m$, where $\hat{p}^{\ast}$ is dimensional pressure. By using parameters from a specific JET deuterium-tritium shot \cite{JETTeam_1992}, we obtain $\rho_m = 2.09 \cdot 10^{-7} \, \text{kg/m}^3$ and the Alfvén velocity $v_{A0} = 5.46 \cdot 10^{6} \, \text{m/s}$. 

\begin{center}
  \includegraphics[width=0.75\columnwidth]{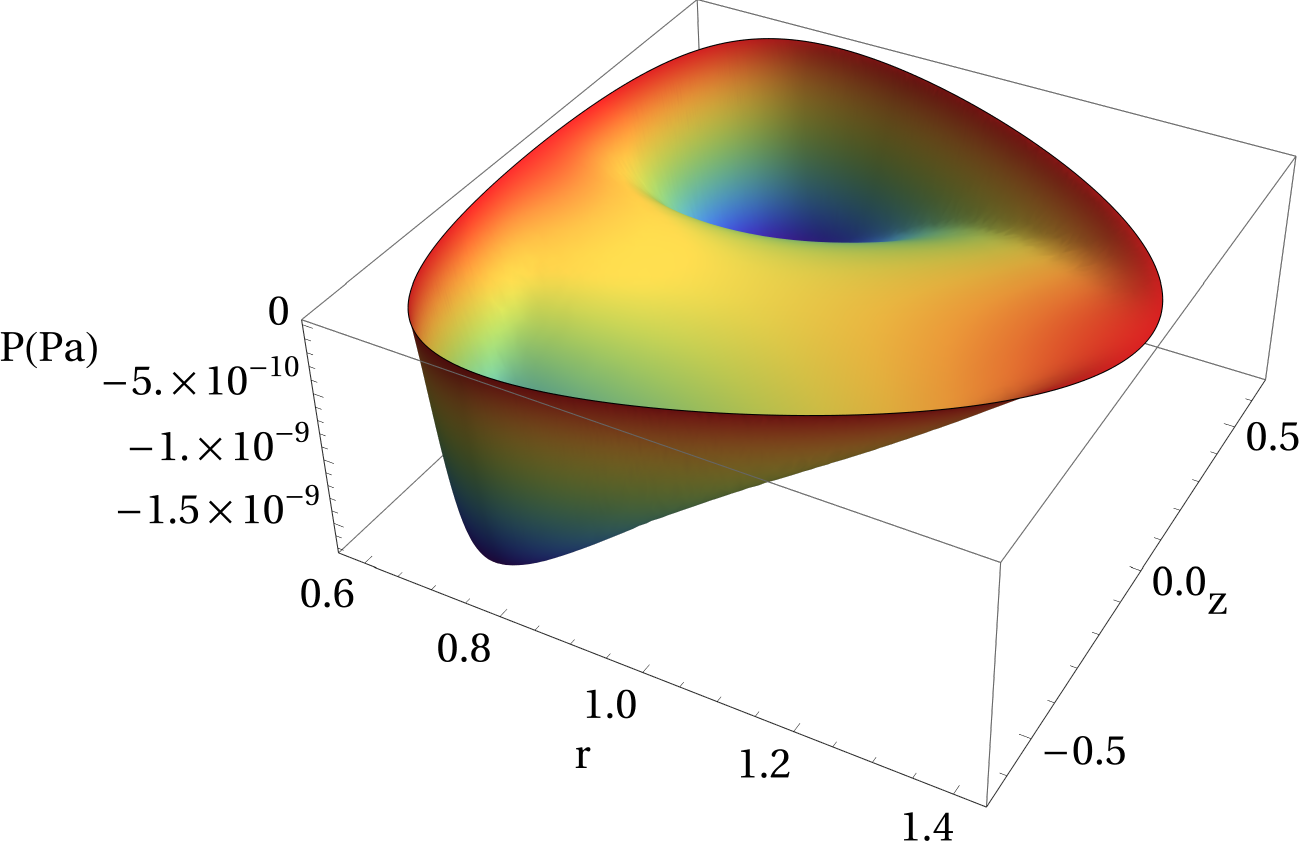}
  \captionof{figure}{Pressure field in Pascal units computed without the application of the drive ($j_D=0$) for a Hartmann number of $H=10^5$.}
  \label{fig: Pressure}
\end{center}

To verify the pressure distribution in the absence of the drive with plasma flow, let us examine Fig.~\ref{fig: Pressure}. The pressure profile is presented in Pascal units (Pa). The order of magnitude of the pressure field in the absence of the current drive turns out to be unrealistically small, as predicted in Section \ref{Pres_grad}.

\begin{center}
  \includegraphics[width=0.75\columnwidth]{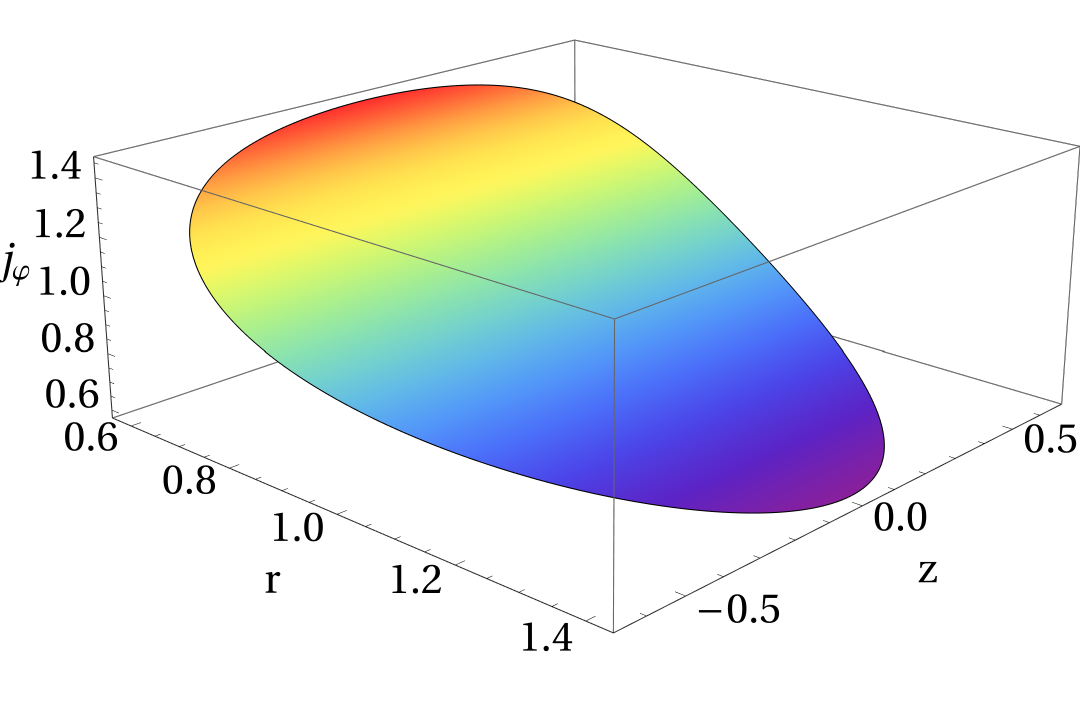} \\
  \vspace{0.5em}  
  \includegraphics[width=0.75\columnwidth]{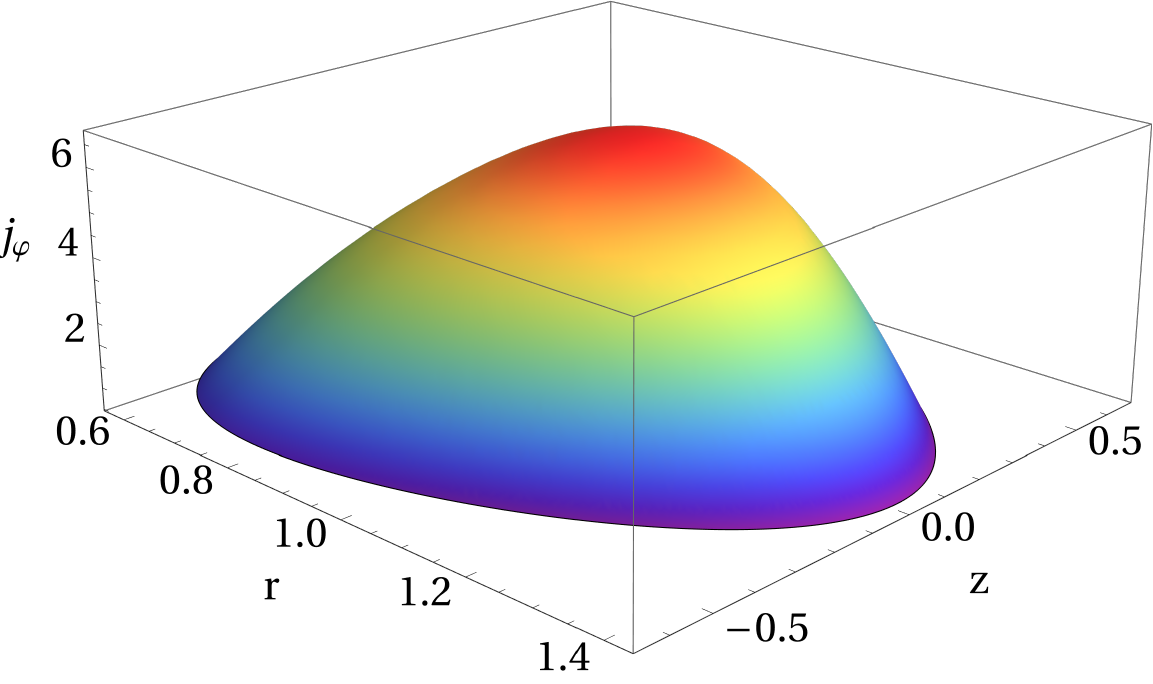}
  \captionof{figure}{Toroidal current field without the drive ($j_D=0$) (top) and with the drive $j_D$ set to $A=100$ and offset $B=0$ (bottom), for a Hartmann number of $H=10$ in dimensionless units.}
  \label{fig: chips}
\end{center}

To explore the effects of a non-inductive current drive, we considered a family of drives, $j_D$, which are solutions to Poisson's equation $\mathbf{\nabla }^{2} j_D = -A$, with the boundary condition $j_D = B$ on $\partial \Omega$. Here, $A$ denotes the magnitude of the drive, while $B$ represents the current distribution offset. This approach provides an initial method for simulating current distributions akin to those generated by heating mechanisms in tokamaks, effectively "adding a bump" to the current profile. \answer{Let us also note here that, by introducing this form of the drive in Eq.~(\ref{drive}), $ E_0/\eta $  start to plays the same role as an offset $B$ as it is a constant. Therefore, in the presence of a current drive, we can set $E_0 = 0$ without loss of generality, as this simply shifts $B$ to $B - E_0/\eta$. This amounts to a purely non-inductive tokamak operation.}

 Let us now choose a drive that produces realistic pressure profiles. To do so, we select $j_D$ with $A = 100$ and $B = 0$. Fig.~\ref{fig: chips} compares the toroidal current density fields, calculated at Hartmann number $H = 10$, using Eqs. (\ref{eq:set_equation1})--(\ref{eq:set_equation5}) for $j_D = 0$ (the reference case) and for $j_D$ with $A = 100$ and $B = 0$. In the original system \cite{kamp2003toroidal, KAMP_MONTGOMERY_2004, oueslati2019numerical,KRUPKA_scaling} ($j_D = 0$), the model fails to produce realistic toroidal current density profiles despite yielding a realistic total current. This is because a ratio of $E_{0}/\eta$ of approximately one corresponds to a realistic total current for the JET deuterium-tritium shot described in \cite{JETTeam_1992}.
\begin{center}
    \includegraphics[width=0.8\columnwidth]{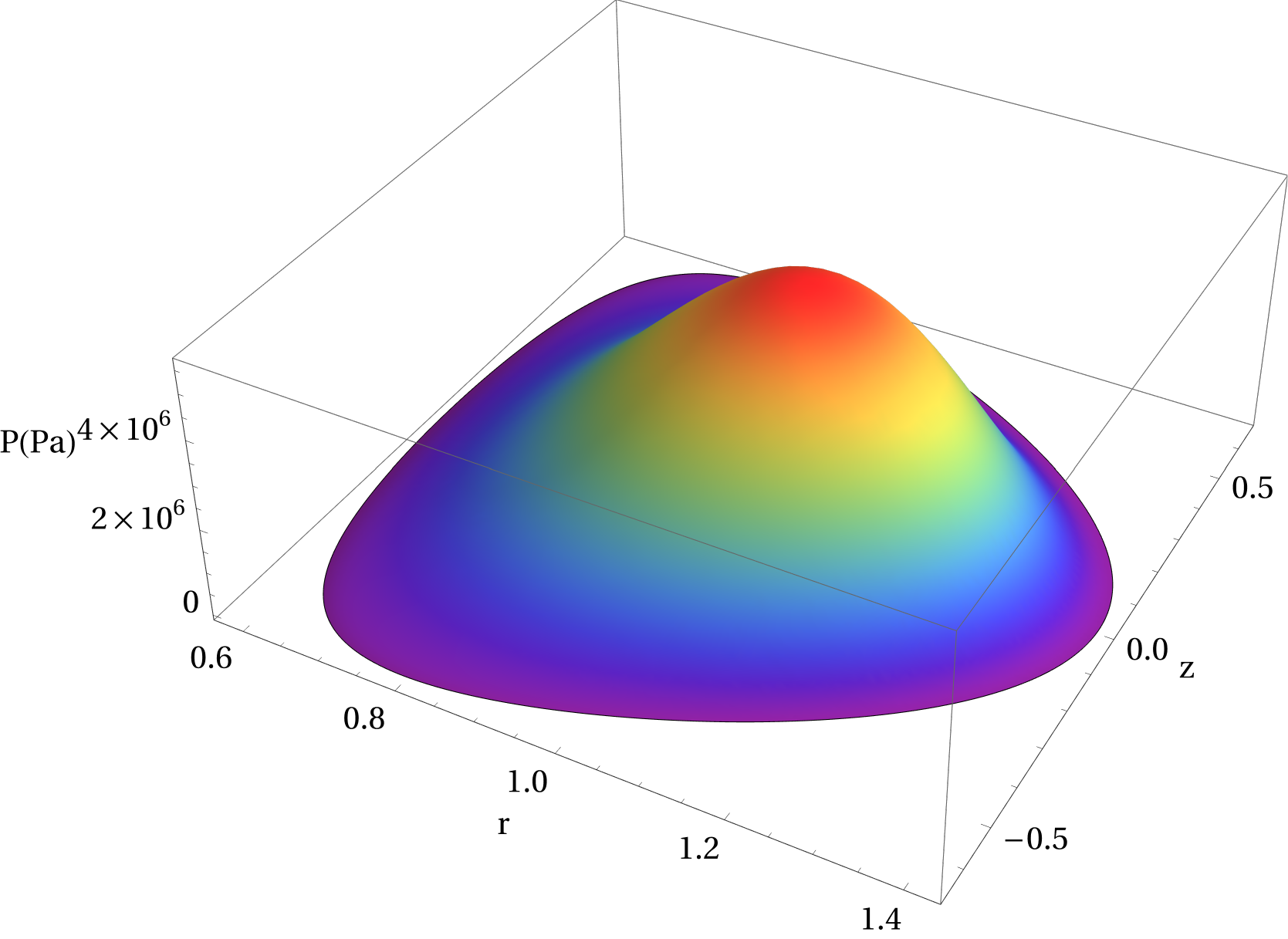}
    \captionof{figure}{Pressure field in Pascal units computed with the application of the drive $j_D$ with $A=100$ and $B=0$ on the toroidal current density field. The Hartmann number is $H=10^5$.}
    \label{fig: PressureDrive}
\end{center}
Let us now examine how the application of the drive affects the pressure profiles. Fig.~\ref{fig: PressureDrive} shows the computed pressure, $p = p^{\ast} - v^2/2$, with the application of the drive $j_D$ using $A = 100$ and $B = 0$ for $H = 10^5$. The current drive not only prevents unrealistically low pressure levels but also establishes a realistic central pressure. In our incompressible framework, what matters physically is the pressure gradient or difference, since one can always add a constant to the pressure field without affecting the dynamics. In our case, the boundary pressure is fixed at zero, which makes the center-to-edge pressure drop a meaningful quantity. For the presented configuration, the central pressure reaches values around $ 4 \times 10^6 $~Pa, in line with experimental observations in the JET tokamak by~\cite{JETTeam_1992}. This is consistent with the observations in ~\cite{Chanine2018}, where the role of absolute pressure values is discussed in the context of incompressible MHD. As pointed out there, pressure differences, and not the absolute value, are what influence the dynamics, especially when interpreting numerical data or comparing with realistic high-$\beta$ experiments. Our results, in this light, confirm that the application of the drive produces physically relevant pressure gradients and magnitudes that fall within realistic tokamak operating regimes.

Let us now examine how the variation of the magnitude of the drive affects the root-mean-square of the pressure as a function of the Hartmann number. Fig.~\ref{fig: Pressure_RMS} illustrates this relationship. It is evident that all the magnitudes $A$ of the drive $j_D$ chosen in this study produce a realistic pressure response.
\begin{center}
    \includegraphics[width=0.8\columnwidth]{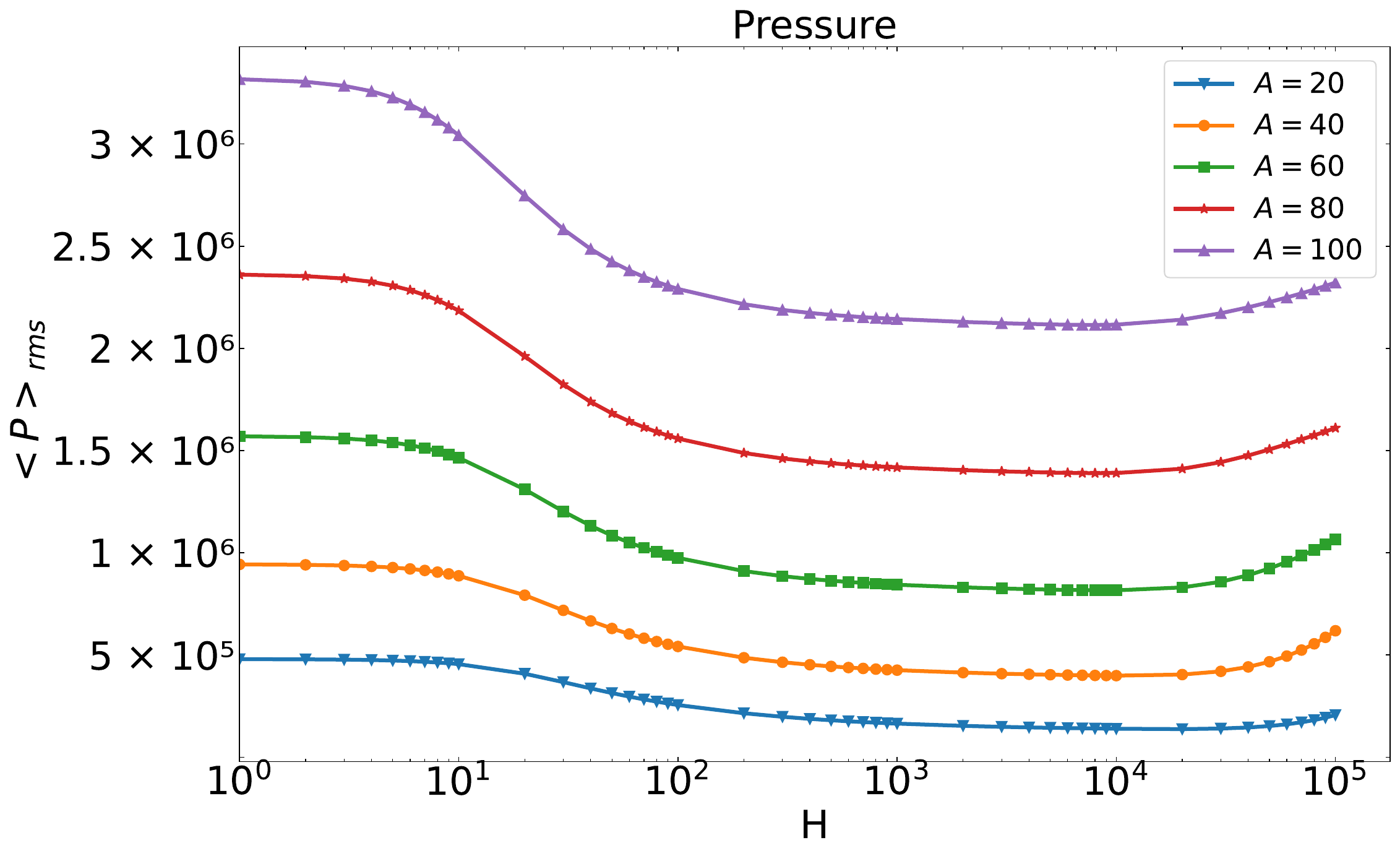}
    \captionof{figure}{Root mean square of the pressure field in pascals as a function of the Hartmann number, with the application of the drive $j_D$ with $B=0$ on the toroidal current field, for various values of $A$.}
    \label{fig: Pressure_RMS}
\end{center}

Let us note that, with the application of the drives, it is possible to achieve various configurations of magnetic flux surfaces, including non-nested magnetic field lines with several $n=0$ islands present. Such an example is given in Fig.~\ref{fig: Non-nested} that shows the magnetic flux surfaces and the pressure profile when the drive $j_D$ with $A=100$ and $B=-5$ is applied. However, we will primarily focus on drives that induce standard nested magnetic flux surfaces.
\begin{center}
    \includegraphics[height=0.55\columnwidth]{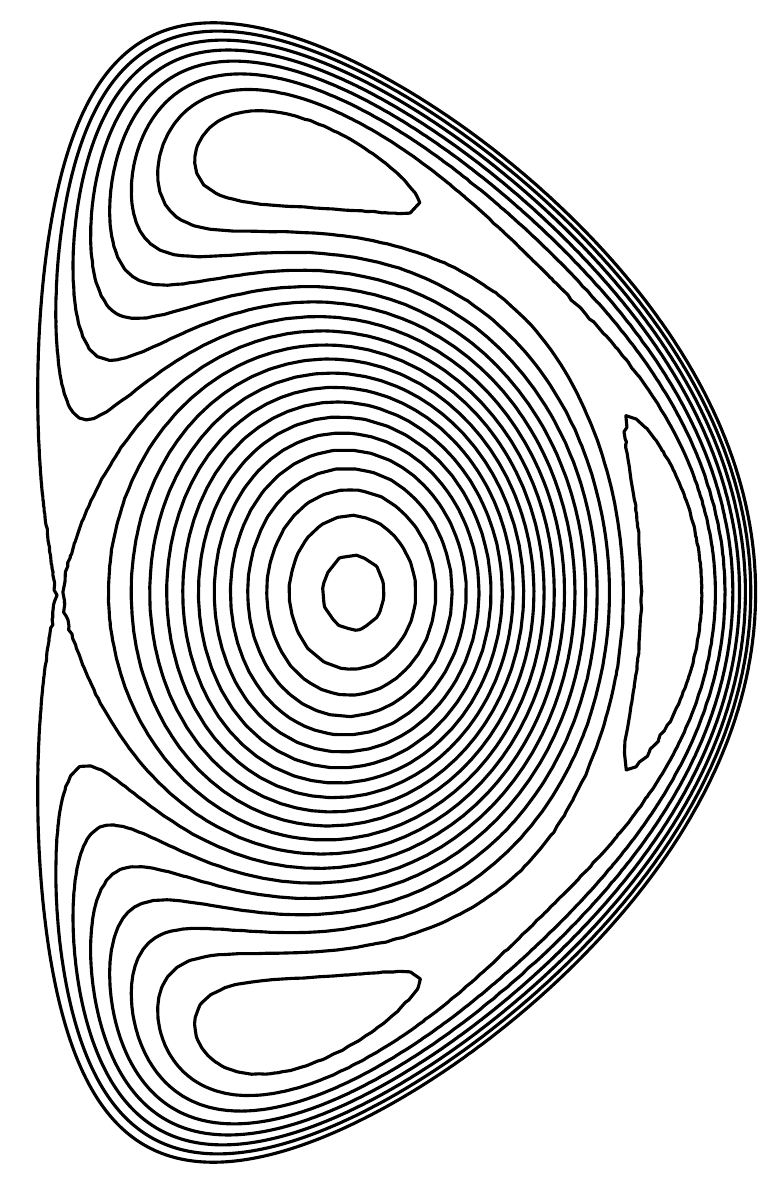}
    \hfill
    \includegraphics[height=0.55\columnwidth]{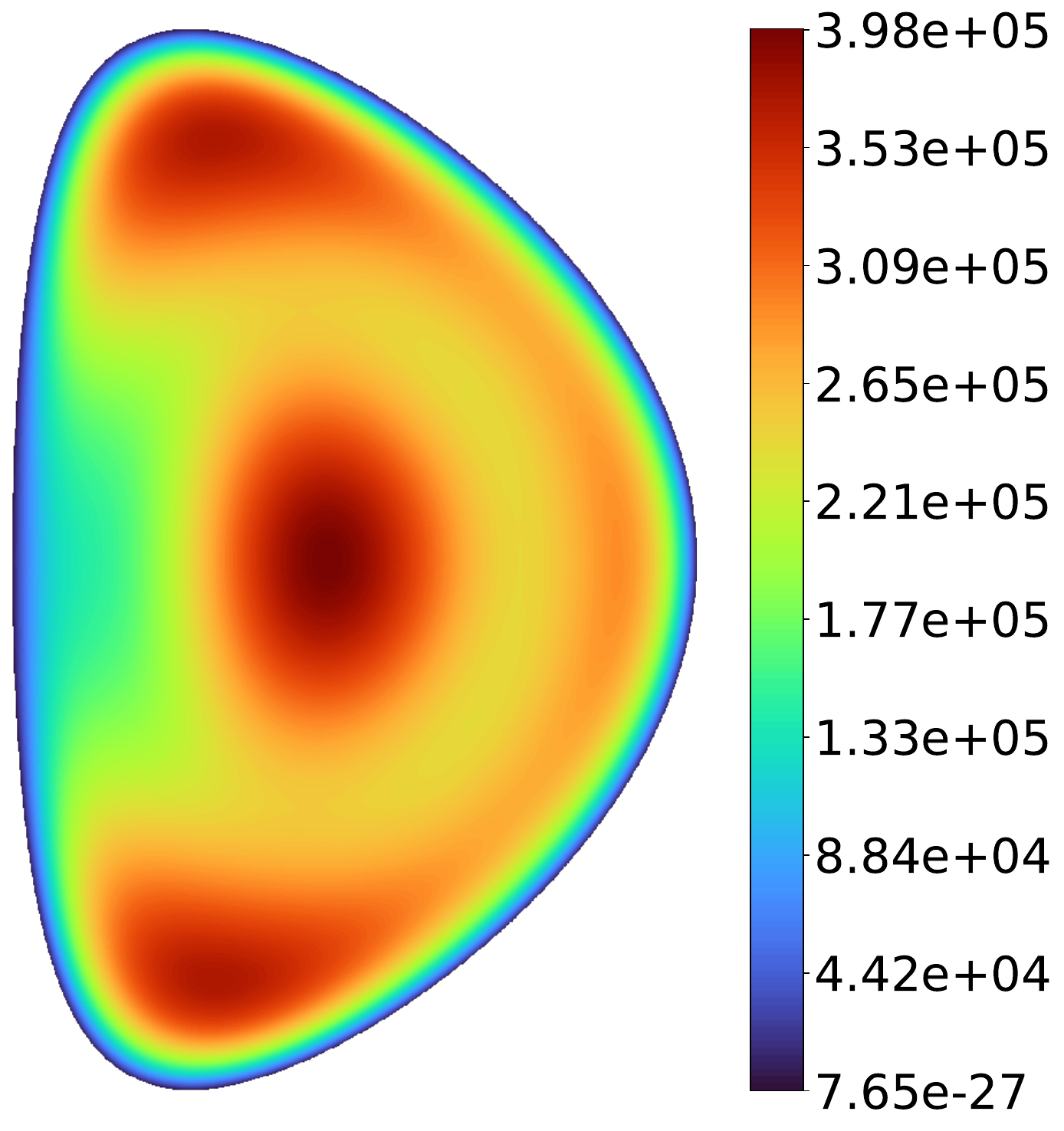}
    \captionof{figure}{Magnetic flux surfaces with internal separatrices (on the left) and pressure profiles (on the right) computed with the application of the drive $j_D$ with $A=100$ and $B=-5$ on the toroidal current field for $H=10^5$.}
    \label{fig: Non-nested}
\end{center}

\subsection{Recovery of Grad–Shafranov equilibrium in the ideal MHD limit}

We now consider the case of standard nested magnetic flux surfaces, induced by a non-inductive current drive $j_D$ with magnitude $A = 100$ and offset $B = 0$. Let us first examine the non-ideal regime, where the dimensionless viscosity $\nu$ and resistivity $\eta$ are both set to 1. This corresponds to a Hartmann number of $H = 1$, and represents a strongly dissipative case with significant viscous and resistive effects. In this regime, shown on the left of Fig.~\ref{fig: Isolines}, the pressure isolines do not align with the magnetic flux surfaces. As we approach the ideal MHD limit, i.e., as $\nu \to 0$ and $\eta \to 0$ (or equivalently, $H \to \infty$), the dynamics become dominated by the ideal force balance. In this case, shown on the right of Fig.~\ref{fig: Isolines} for $H = 10^3$, the pressure isolines start to align with the magnetic flux surfaces. This result shows that the pressure tends to become a flux function in the ideal limit.
\begin{center}
    \includegraphics[height=0.6\columnwidth]{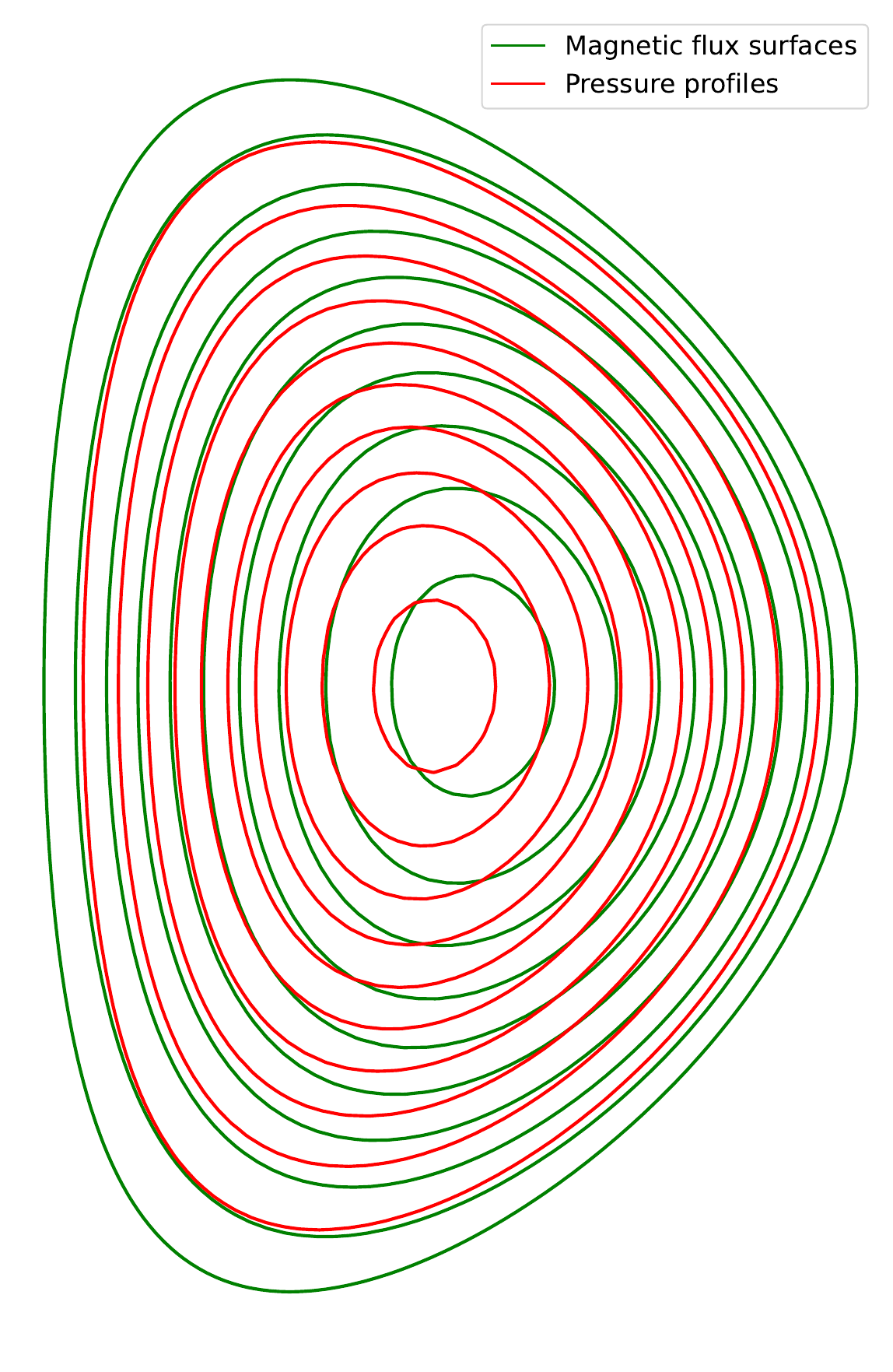}
   \qquad\qquad
    \includegraphics[height=0.6\columnwidth]{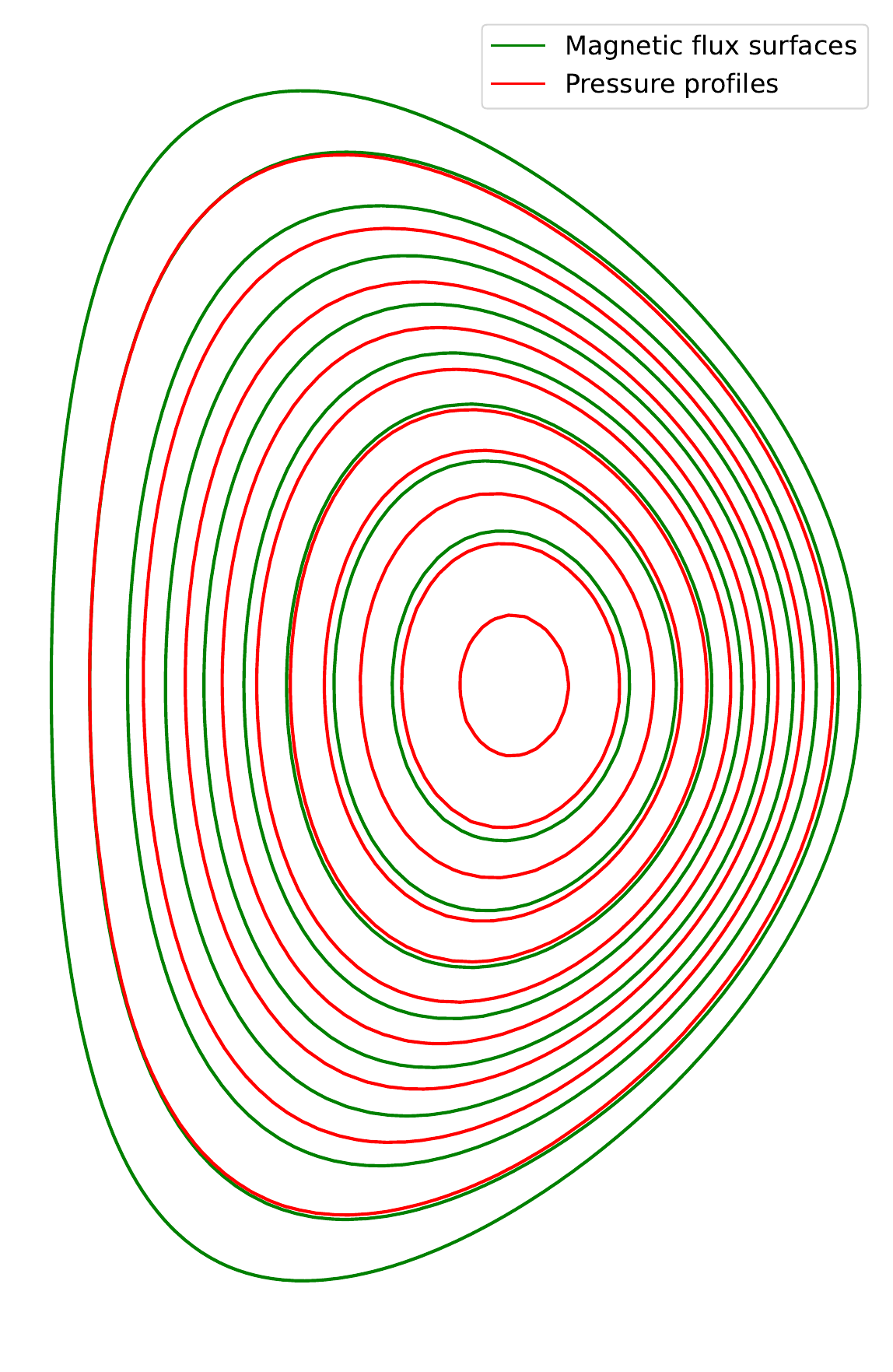}
    \captionof{figure}{Magnetic flux surfaces and pressure isolines computed for the drive $j_D$ with $A = 100$ and $B = 0$ with $H = 1$ (non-ideal case) on the left and  $H = 10^3$ (approaching the ideal limit) on the right.}
    \label{fig: Isolines}
\end{center}

To explore this further, we extract the data for the pressure field $P$ and the diamagnetic function $F = x B_\varphi$, and plot them as functions of the corresponding values of the poloidal magnetic flux, $\psi$. In the left-hand plots of Figure~\ref{fig: comparison_MHDGS}, we show that for $H = 1000$, both $P$ and $F$ are very well approximated by cubic polynomial functions of $\psi$. They behave then as flux functions. Then we use these fits in a code that solves the Grad–Shafranov equation using a Picard fixed-point iteration.

The right-hand plot of Figure~\ref{fig: comparison_MHDGS} compares the level curves of the normalized poloidal magnetic flux, $\psi_N$, obtained for the same $\psi_N$ values from a self-consistent visco-resistive MHD code at $H = 1000$, with those obtained from the Grad–Shafranov equilibrium reconstruction. A very good agreement is observed, as expected, since the non-ideal and finite-velocity contributions are small. Quantitatively, using the $L^2$-norm 
\begin{equation}
 \left\| \psi_1 - \psi_2 \right\|_{L^2(\Omega)} = \left( \int_\Omega \left( \psi_1(x, y) - \psi_2(x, y) \right)^2 \, dx\,dy \right)^{1/2},  
\end{equation}
we get in this case $\left\| \psi_{N,GS} - \psi_{N,MHD} \right\|_{L^2(\Omega)} \approx 0.015$.

    \begin{center}
      \includegraphics[height=0.6\columnwidth]{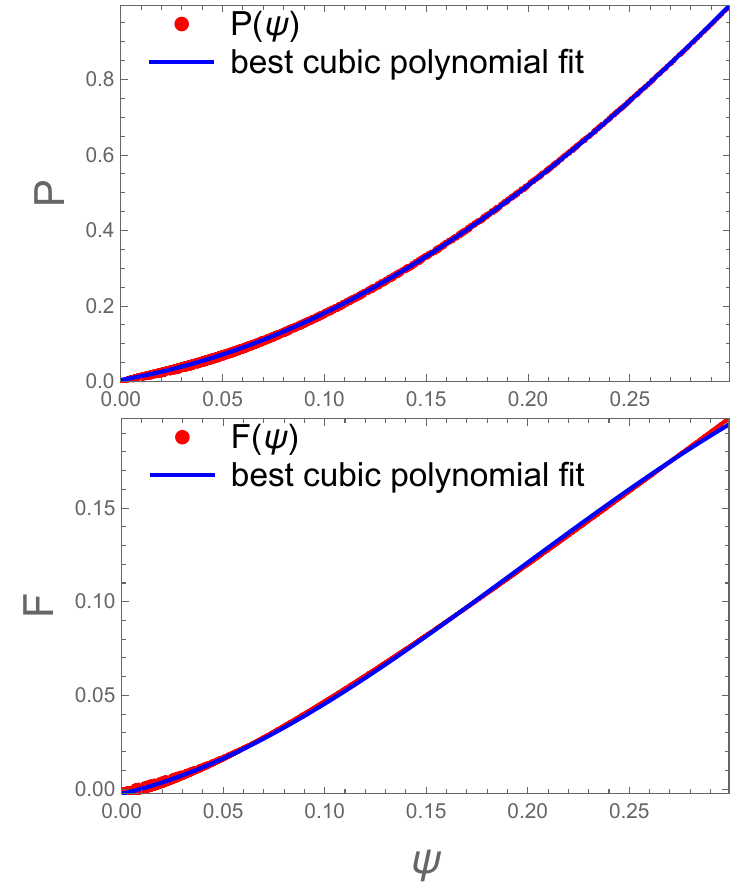} 
   \hfill
      \includegraphics[height=0.6\columnwidth]{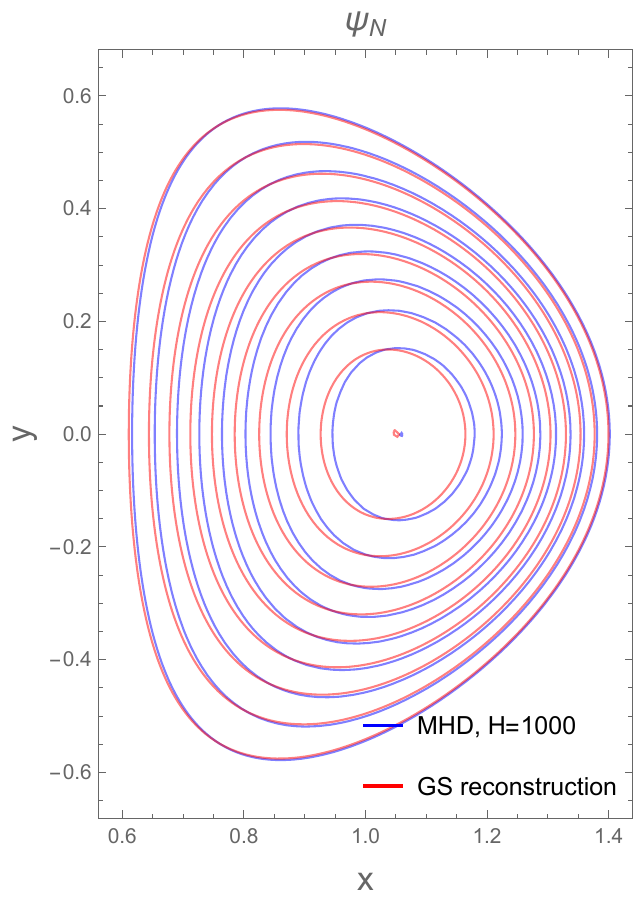}
    \captionof{figure}{(Left) Dimensionless pressure and diamagnetic function plotted as functions of the poloidal magnetic flux in the $H = 1000$ MHD simulation, along with their best cubic polynomial fits. The $(\psi, P)$ and $(\psi, F)$ plots are constructed by evaluating $P$, $F$, and $\psi$ at the same finite element mesh nodes $(x_i, y_i)$. Each pair $(\psi_i, P_i)$ and $(\psi_i, F_i)$ is represented as a red point. (Right) Comparison of the same levels of the normalized poloidal magnetic flux, $\psi_N$, obtained from the $H = 1000$ visco-resistive MHD simulation (blue curves) and from the Grad–Shafranov equilibrium reconstruction (red curves) using the pressure and diamagnetic functions fitted in the left panel.}
    \label{fig: comparison_MHDGS}
\end{center}

This indicates that the Grad-Shafranov equilibrium can be recovered from the self-consistent MHD solutions in the appropriate limits, but this also highlights the broader validity of our model in more dissipative regimes. Most importantly, the pressure field is self-consistently determined within our system, which is not the case for Grad-Shafranov equilibrium reconstructions.

\subsection{Impact of the non-inductive current drive on steady-state velocity and scaling}

Let us now examine the impact on the velocity distribution of the application of the current drive $j_D$. Fig.~\ref{fig: Lapl_A} presents the root-mean-square of the toroidal velocity field while applying the drive $j_D$ with $B = 0$ to the toroidal current field across various values of $A$, as in Fig.~\ref{fig: Pressure_RMS}. It can be observed that while varying the magnitude of the drive causes an increase in velocities in the low-$H$ regime, the large-$H$, boundary layer regime remains almost unchanged, despite the application of current drives with different magnitudes.

In \cite{KRUPKA_scaling}, we showed that at large Hartmann numbers, the velocity field develops a distinct boundary layer that becomes progressively thinner as $H$ increases. In this regime, the boundary layer thickness and, consequently, the velocity, scale with the Hartmann number, if non-linear effects remain negligible. Specifically, by analyzing the boundary layer equations analytically, we found that both the toroidal and poloidal velocities scale as $H^{1/4}$ for $H \gg 1$. This scaling was also confirmed numerically through power-law fitting. Increasing the magnitude of the drive raises the total current, but the velocities appear unaffected by this variation. This insensitivity is consistent with the persistence of the boundary layer regime, which imposes a fixed scaling of the velocities with the Hartmann number, independent of the drive amplitude.

\begin{center}
    \includegraphics[width=0.9\columnwidth]{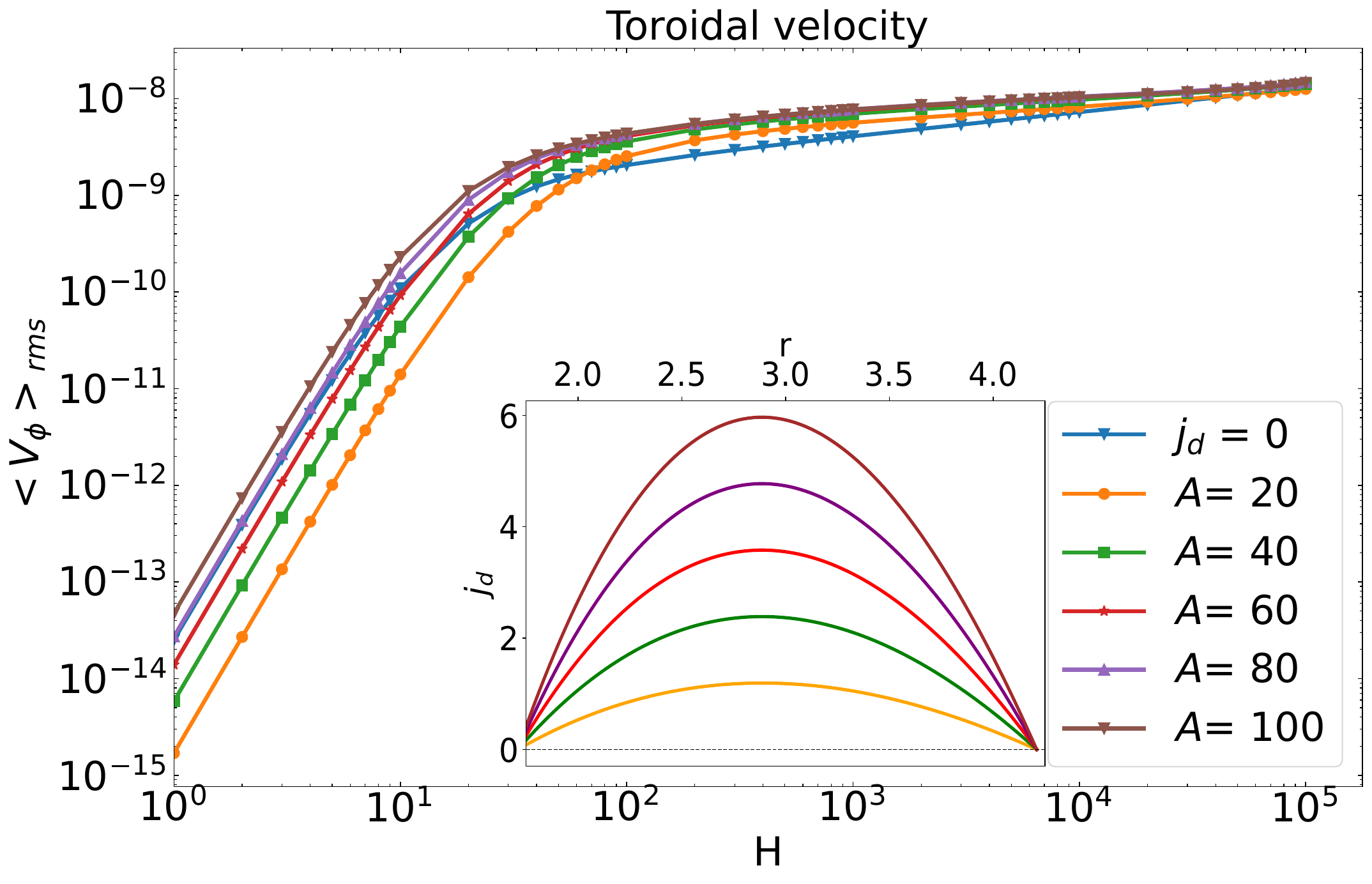}
    \captionof{figure}{Root mean square of the toroidal velocity in Alfvén velocity units as a function of the Hartmann number, considering the application of the drive $j_D$ with $B=0$ on the toroidal current field, for the different values of $A$.}
    \label{fig: Lapl_A}
\end{center}

\begin{center}    
    \includegraphics[width=0.9\columnwidth]{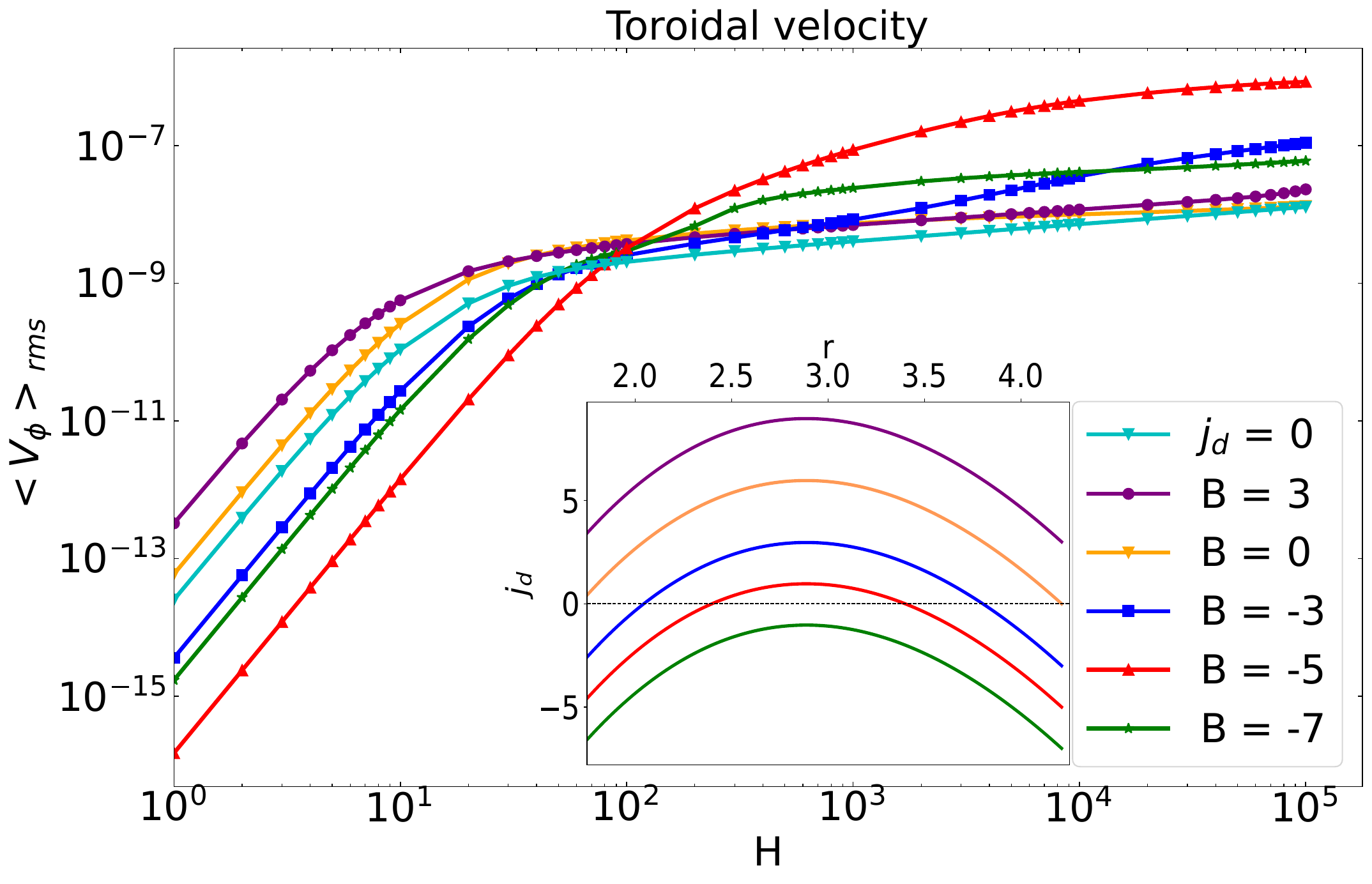}
    \captionof{figure}{Root mean square of the toroidal velocity in Alfvén velocity units as a function of the Hartmann number, considering the application of the drive $j_D$ with $A=100$ on the toroidal current field, for various values of $B$.}
    \label{fig: Lapl_B}
\end{center}

Next, let us examine how the velocities change with the variation of the parameter $B$, which represents the offset of the drive $j_D$. Fig.~\ref{fig: Lapl_B} presents the same information as Fig.~\ref{fig: Lapl_A}, but with a fixed value of $A=100$ while exploring different values of $B$. Shifting the drive results in the highest velocities at $B=-5$. The usual large-$H$ velocity behaviour changes at certain parameters of the current drive. 
Here we consider several different toroidal current profiles: two that lie entirely in the positive region (e.g. $B=3$ and $B=0$), two that are in between positive and negative regions ($B=-3$ and $B=-5$), and one that is fully in the negative region $B=-7$. In some of our simulations, the self-consistent MHD evolution leads to the formation of reversed current profiles, where the toroidal current density changes sign across the plasma radius. These current configurations give rise to additional magnetic axes and separatrices, modifying the magnetic topology beyond the typical single-axis case. Such phenomena are not only robust numerical features but are also physically relevant, as reversed current profiles and associated topological changes have been experimentally observed in devices like JET and JT-60U. Moreover, we found that these configurations coincide with a breakdown of standard scaling laws for the velocity fields, showing increased root-mean-square values for both toroidal and poloidal velocities. This makes them particularly interesting for understanding fundamental transport and equilibrium behaviour in such regimes.

\begin{center}
    \includegraphics[height=0.525\columnwidth]{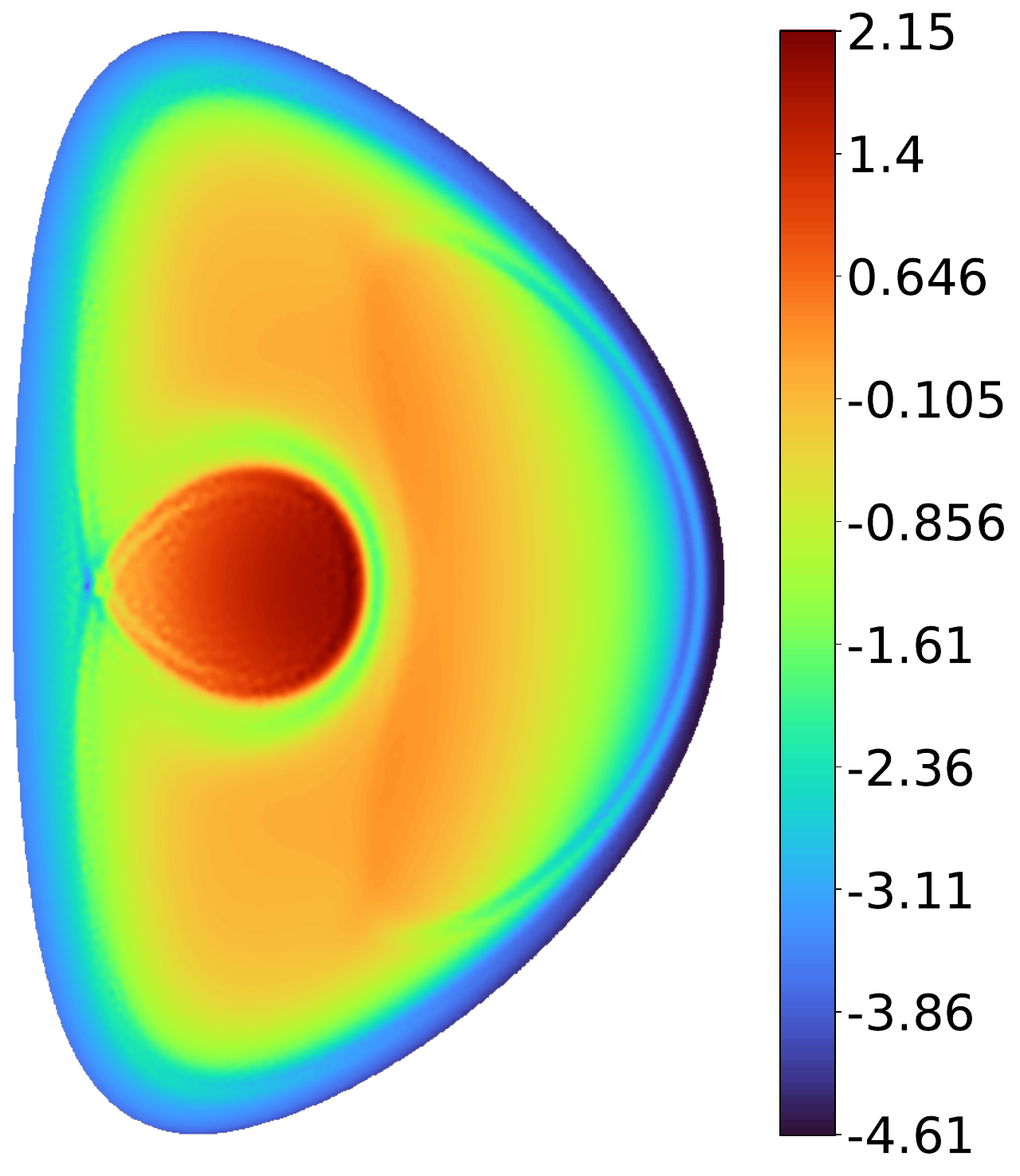}
    \hfill
    \includegraphics[height=0.525\columnwidth]{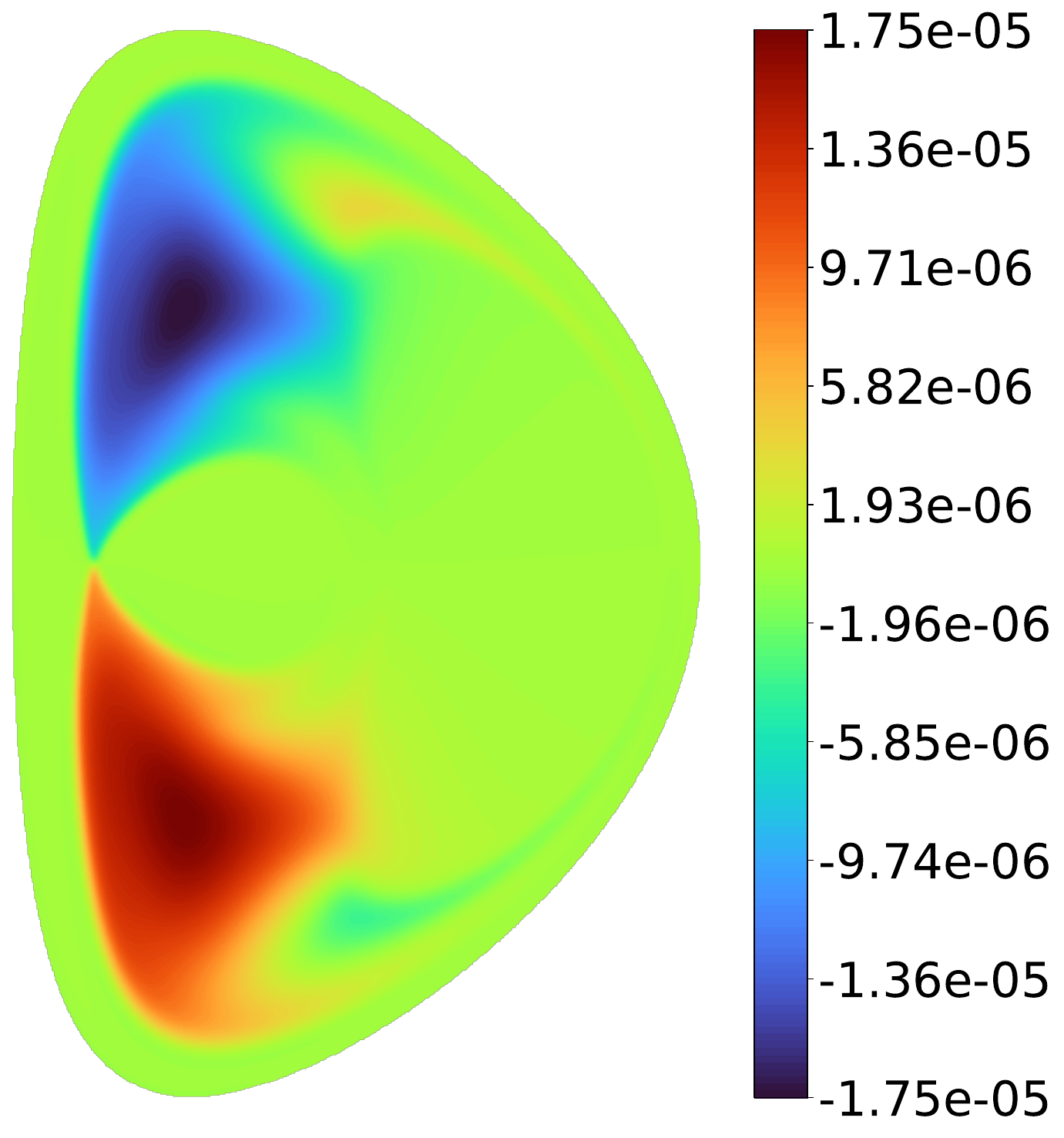}
    \captionof{figure}{Toroidal current density field (on the left) and toroidal velocity ﬁeld (on the right) with the application of the drive $j_D$ with $A=100$, $B=-5$ for $H=10^5$ as in Fig. \ref{fig: Non-nested}.}
    \label{fig: drive change}
\end{center}

Indeed, in the high-Hartmann number regime, two scenarios emerge: either the boundary layer forms~\cite{KRUPKA_scaling}, and the root-mean-square of the toroidal and poloidal velocities exhibit some scaling law with the Hartmann number, as is the case for the drive $j_D$ with $B=0$ and $B=3$; or the velocities do not develop a linear behavior on a log-log scale for $B=-3$, $-5$, and $-7$, meaning that the velocities do not scale with the Hartmann number. This departure from scaling behavior is probably linked to the way the toroidal velocity field changes with the drive offset $B$. For different values of $B$, the velocity distribution can differ significantly, which affects whether or not the boundary layer forms. In our view, what we observe here is a competition between different scaling regimes, which leads to the non-monotonic behavior. Let us now take a closer look at this latest phenomenology, which is novel and pertains to a situation involving non-nested magnetic flux surfaces with internal separatrices. To illustrate this, we examine the non-inductive current drive $j_D$ with $A = 100$ and $B = -5$ for $H = 10^5$, where the magnetic flux surfaces and pressure field are depicted in Fig. \ref{fig: Non-nested}. Fig.~\ref{fig: drive change} shows on the left the associated toroidal current field. 

The velocity distribution on the right of Fig.~\ref{fig: drive change} closely resembles the toroidal current distribution, with the highest velocities occurring at the transition point between positive and negative current regions. In this case, we observe no formation of a boundary layer, which is advantageous from a numerical perspective. The absence of a boundary layer contributes to increased stability in the code and yields more robust results. The toroidal velocity field in our simulations exhibits a clear up-down antisymmetry, such that $ v_\varphi(x, y) = -v_\varphi(x, -y)$, leading to zero net toroidal momentum. This symmetry arises naturally from the geometry and boundary conditions imposed in our setup. While this excludes the possibility of spontaneous momentum generation in the present context, it is worth noting that previous studies have demonstrated mechanisms through which up-down symmetry breaking can lead to the emergence of net toroidal flow, both in gyrokinetic frameworks~\cite{Camenen2009} and in MHD~\cite{Moraler2012}. For instance, it was shown in~\cite{oueslati2020breaking} that breaking this symmetry by means of external magnetic perturbations can influence flow topology and might be used to induce finite mean flows. There is significant potential to achieve much higher velocities with these drives; however, accurately predicting which current drive would be optimal for maximizing velocities remains a challenge and necessitates further investigation. 

\section{\label{Conclusion}Conclusion and perspectives}

We have examined axisymmetric steady states of tokamak plasmas using a self-consistent incompressible visco-resistive MHD model with external drives. While acknowledging the intrinsic limitations of a magnetohydrodynamic (rather than kinetic) approach, this setup remains relevant to the fully non-inductive regimes pursued in advanced fusion reactor designs. The knowledge of axisymmetric steady states is fundamental for stability analysis, but also for a recently proposed classification of magnetic and current density modes \cite{Firpo_2024,Firpo_2025}.

A significant advancement of the present study is the formulation of a Poisson's equation governing the pressure within this self-consistent incompressible visco-resistive MHD framework. This allows for the computation of the pressure profile as soon as its boundary condition is prescribed. This approach differs notably from the Grad-Shafranov method used in equilibrium reconstruction, where pressure is treated as a free function. The numerical solution of this Poisson equation for pressure obtained using the finite element method demonstrates the necessity of implementing additional drives to avoid unrealistic pressure profiles with zero gradients in the ideal and no-flow limit. We have shown that this can be achieved through non-inductive-like current drives, resulting in realistic pressure profiles. 

We examined a family of functions to model the non-inductive current drives of tokamaks, but further research is needed to optimize the distribution of non-inductive currents and make them more realistic. Another goal is to maximize their effectiveness in enhancing plasma speed and achieving fusion-relevant pressure profiles. Notably, certain configurations of our test current drives led to the formation of internal separatrices and non-nested magnetic flux surfaces, indicating the potential for more complex equilibrium structures under specific plasma conditions. An additional outlook of this study is to leverage the supplementary constraint offered by the Poisson equation governing the pressure, with the aim of enhancing or constraining equilibrium reconstruction in steady-state operational regimes. 

Finally, the resulting toroidal current profiles were found to depend on the Hartmann number. This suggests that future implementations of these drives could involve fixed toroidal current density profiles that are independent of other system parameters. This topic will be explored further in subsequent work and is discussed in more detail in \cite{Thesis}.

\section*{Acknowledgments}

This work has been carried out within the framework of the EUROfusion Consortium and has received funding from the Euratom Research and Training Programme 2014-2018 under grant agreement No 633053. The views and opinions expressed herein do not necessarily reflect those of the European Commission.

\section*{Declaration of interests}

The authors report no conflict of interest.

\section*{Data availability statement}
The data supporting the findings of this study are available from the corresponding author upon reasonable request.


\begin{thebibliography}{34}
\expandafter\ifx\csname natexlab\endcsname\relax\def\natexlab#1{#1}\fi
\def\au#1{#1} \def\ed#1{#1} \def\yr#1{#1}\def\at#1{#1}\def\jt#1{\textit{#1}} \def\bt#1{#1}\def\bvol#1{\textbf{#1}} \def\vol#1{#1} \def\pg#1{#1} \def\publ#1{#1}\def\arxiv#1{#1}\def\org#1{#1}\def\st#1{\textit{#1}}

\bibitem[Ball {\em et~al.\/}(2020)Ball, Brunner \& C.J.]{Ball2020}
{\sc \au{Ball, Justin}, \au{Brunner, Stephan} \& \au{C.J., Ajay}} \yr{2020}  \at{Eliminating turbulent self-interaction through the parallel boundary condition in local gyrokinetic simulations}.  \jt{Journal of Plasma Physics}  \bvol{86}~(2),  \pg{905860207}.

\bibitem[Braams(1991)]{Braams1991}
{\sc \au{Braams, B~J}} \yr{1991}  \at{The interpretation of tokamak magnetic diagnostics}.  \jt{Plasma Physics and Controlled Fusion}  \bvol{33}~(7),  \pg{715}.

\bibitem[Camenen {\em et~al.\/}(2009)Camenen, Peeters, Angioni, Casson, Hornsby, Snodin \& Strintzi]{Camenen2009}
{\sc \au{Camenen, Y.}, \au{Peeters, A.~G.}, \au{Angioni, C.}, \au{Casson, F.~J.}, \au{Hornsby, W.~A.}, \au{Snodin, A.~P.} \& \au{Strintzi, D.}} \yr{2009}  \at{Intrinsic rotation driven by the electrostatic turbulence in up-down asymmetric toroidal plasmas}.  \jt{Physics of Plasmas}  \bvol{16}~(6),  \pg{062501},  \arxiv{arXiv: https://pubs.aip.org/aip/pop/article-pdf/doi/10.1063/1.3138747/15843760/062501\_1\_online.pdf}.

\bibitem[Cappello \& Escande(2000)]{Cappello}
{\sc \au{Cappello, S.} \& \au{Escande, D.~F.}} \yr{2000}  \at{Bifurcation in {{Viscoresistive MHD}}: {{The Hartmann Number}} and the {{Reversed Field Pinch}}}.  \jt{Physical Review Letters}  \bvol{85}~(18),  \pg{3838--3841}.

\bibitem[Chahine \& Bos(2018)]{Chanine2018}
{\sc \au{Chahine, Robert} \& \au{Bos, Wouter J.~T.}} \yr{2018}  \at{On the role and value of beta in incompressible {MHD} simulations}.  \jt{Physics of Plasmas}  \bvol{25}~(4),  \pg{042115},  \arxiv{arXiv: https://pubs.aip.org/aip/pop/article-pdf/doi/10.1063/1.5018666/13774613/042115\_1\_online.pdf}.

\bibitem[Daza {\em et~al.\/}(2024)Daza, Reynolds-Barredo, Sanchez, Loarte \& Tribaldos]{Torija_Daza_2024}
{\sc \au{Daza, G. F.-Torija}, \au{Reynolds-Barredo, J.M.}, \au{Sanchez, R.}, \au{Loarte, A.} \& \au{Tribaldos, V.}} \yr{2024}  \at{Flipec, an ideal {MHD} free-boundary axisymmetric equilibrium solver in the presence of macroscopic flows}.  \jt{Nuclear Fusion}  \bvol{64}~(8),  \pg{086012}.

\bibitem[Del~Prete \& Montani(2021)]{DelPrete2021}
{\sc \au{Del~Prete, Matteo} \& \au{Montani, Giovanni}} \yr{2021}  \at{Influence of rotation on axisymmetric plasma equilibria: double-null {DTT} scenario}.  \jt{Plasma Physics and Controlled Fusion}  \bvol{63}~(12).

\bibitem[Ferron {\em et~al.\/}(2013)Ferron, Holcomb, Luce, Park, Politzer, Turco, Heidbrink, Doyle, Hanson, Hyatt, In, La~Haye, Lanctot, Okabayashi, Petrie, Petty \& Zeng]{Ferron_2013}
{\sc \au{Ferron, J.~R.}, \au{Holcomb, C.~T.}, \au{Luce, T.~C.}, \au{Park, J.~M.}, \au{Politzer, P.~A.}, \au{Turco, F.}, \au{Heidbrink, W.~W.}, \au{Doyle, E.~J.}, \au{Hanson, J.~M.}, \au{Hyatt, A.~W.}, \au{In, Y.}, \au{La~Haye, R.~J.}, \au{Lanctot, M.~J.}, \au{Okabayashi, M.}, \au{Petrie, T.~W.}, \au{Petty, C.~C.} \& \au{Zeng, L.}} \yr{2013}  \at{Progress toward fully noninductive discharge operation in {DIII-D} using off-axis neutral beam injection}.  \jt{Physics of Plasmas}  \bvol{20}~(9),  \pg{092504},  \arxiv{arXiv: https://pubs.aip.org/aip/pop/article-pdf/doi/10.1063/1.4821072/15767437/092504\_1\_online.pdf}.

\bibitem[Firpo(2024)]{Firpo_2024}
{\sc \au{Firpo, M.-C.}} \yr{2024}  \at{Interplay of the magnetic and current density field topologies in axisymmetric devices for magnetic confinement fusion}.  \jt{Journal of Plasma Physics}  \bvol{90}~(6),  \pg{175900601}.

\bibitem[Firpo(2025)]{Firpo_2025}
{\sc \au{Firpo, M.-C.}} \yr{2025}  \at{Cross-analysis of magnetic and current density field topologies in a {Quiescent High Confinement Mode} tokamak discharge}.  \jt{Foundations}  \bvol{5}~(2).

\bibitem[Garzotti {\em et~al.\/}(2018)Garzotti, Barbato, Garcia, Hayashi, Voitsekhovitch, Giruzzi, Maget, Romanelli, Saarelma, Stankiewitz, Yoshida \& Zagórski]{Garzotti_2018}
{\sc \au{Garzotti, L.}, \au{Barbato, E.}, \au{Garcia, J.}, \au{Hayashi, N.}, \au{Voitsekhovitch, I.}, \au{Giruzzi, G.}, \au{Maget, P.}, \au{Romanelli, M.}, \au{Saarelma, S.}, \au{Stankiewitz, R.}, \au{Yoshida, M.} \& \au{Zagórski, R.}} \yr{2018}  \at{Analysis of {JT-60SA} operational scenarios}.  \jt{Nuclear Fusion}  \bvol{58}~(2),  \pg{026029}.

\bibitem[Guazzotto {\em et~al.\/}(2004)Guazzotto, Betti, Manickam \& Kaye]{Guazzotto2004}
{\sc \au{Guazzotto, L.}, \au{Betti, R.}, \au{Manickam, J.} \& \au{Kaye, S.}} \yr{2004}  \at{Numerical study of tokamak equilibria with arbitrary flow}.  \jt{Physics of Plasmas}  \bvol{11}~(2),  \pg{604--614},  \arxiv{arXiv: https://pubs.aip.org/aip/pop/article-pdf/11/2/604/19177034/604{\textbackslash}\_1{\textbackslash}\_online.pdf}.

\bibitem[Hecht(2012)]{hecht2012new}
{\sc \au{Hecht, Frédéric}} \yr{2012}  \at{New development in {{FreeFem}}++}.  \jt{Journal of Numerical Mathematics}  \bvol{20}~(3-4),  \pg{251--265}.

\bibitem[van Houtte {\em et~al.\/}(2004)van Houtte, Martin, Bécoulet, Bucalossi, Giruzzi, Hoang, Loarer, Saoutic \& (on behalf of~the Tore Supra~Team)]{Houtte_2004}
{\sc \au{van Houtte, D.}, \au{Martin, G.}, \au{Bécoulet, A.}, \au{Bucalossi, J.}, \au{Giruzzi, G.}, \au{Hoang, G.T.}, \au{Loarer, Th.}, \au{Saoutic, B.} \& \au{(on behalf of~the Tore Supra~Team)}} \yr{2004}  \at{Recent fully non-inductive operation results in {Tore Supra} with 6 min, 1 {GJ} plasma discharges}.  \jt{Nuclear Fusion}  \bvol{44}~(5),  \pg{L11}.

\bibitem[{JET Team}(1992)]{JETTeam_1992}
{\sc \au{{JET Team}}} \yr{1992}  \at{Fusion energy production from a deuterium-tritium plasma in the {JET} tokamak}.  \jt{Nuclear Fusion}  \bvol{32}~(2),  \pg{187}.

\bibitem[Kaltsas \& Throumoulopoulos(2022)]{Kaltsas2022}
{\sc \au{Kaltsas, D.~A.} \& \au{Throumoulopoulos, G.~N.}} \yr{2022}  \at{Neural network tokamak equilibria with incompressible flows}.  \jt{Physics of Plasmas}  \bvol{29}~(2).

\bibitem[Kamp \& Montgomery(2003)]{kamp2003toroidal}
{\sc \au{Kamp, L.~P.} \& \au{Montgomery, D.~C.}} \yr{2003}  \at{Toroidal flows in resistive magnetohydrodynamic steady states}.  \jt{Physics of Plasmas}  \bvol{10},  \pg{157--167}.

\bibitem[Kamp \& Montgomery(2004)]{KAMP_MONTGOMERY_2004}
{\sc \au{Kamp, Leon~P.} \& \au{Montgomery, David~C.}} \yr{2004}  \at{Toroidal steady states in visco-resistive magnetohydrodynamics}.  \jt{Journal of Plasma Physics}  \bvol{70}~(2),  \pg{113--142}.

\bibitem[Kamp {\em et~al.\/}(1998)Kamp, Montgomery \& Bates]{Kamp1998}
{\sc \au{Kamp, Leon~P.}, \au{Montgomery, David~C.} \& \au{Bates, Jason~W.}} \yr{1998}  \at{Toroidal flows in resistive magnetohydrodynamic steady states}.  \jt{Physics of Fluids}  \bvol{10}~(7),  \pg{1757--1766},  \arxiv{arXiv: https://pubs.aip.org/aip/pof/article-pdf/10/7/1757/12487686/1757{\textbackslash}\_1{\textbackslash}\_online.pdf}.

\bibitem[Kikuchi(2010)]{Kikuchi2010}
{\sc \au{Kikuchi, Mitsuru}} \yr{2010}  \at{A review of fusion and tokamak research towards steady-state operation: A {JAEA} contribution}.  \jt{Energies}  \bvol{3}~(11),  \pg{1741--1789}.

\bibitem[Ko {\em et~al.\/}(2024)Ko, Yoon, Kim, Kwak, Park, Nam, Wang, Chung, Park, Park, Lee, Han, Choi, Na, In, Lee, Kim, Yun, Ghim, Choe, Kwon, Lee, Lee, Jeon, Kim, Lee, Shin, Kim, Lee, Hahn, Lee, Kim, Bak, Lee, Lee, Jeong, Woo, Kim, Juhn, Ko, Sung, Shin, Park, Kim, Park, Logan, Yang, Kolemen, Hu, Shousha, Barr, Paz-Soldan, Park, Sabbagh, Ida, Kim, Loarte, Gilson, Eldon, Nakano, Tala \& Team]{Ko_2024}
{\sc \au{Ko, Won-Ha}, \au{Yoon, S.W.}, \au{Kim, W.C.}, \au{Kwak, J.G.}, \au{Park, K.L.}, \au{Nam, Y.U.}, \au{Wang, S.J.}, \au{Chung, J.}, \au{Park, B.H.}, \au{Park, G.Y.}, \au{Lee, H.H.}, \au{Han, H.S.}, \au{Choi, M.J.}, \au{Na, Y.S.}, \au{In, Y.}, \au{Lee, C.Y.}, \au{Kim, M.}, \au{Yun, G.S.}, \au{Ghim, Y.-C.}, \au{Choe, W.H.}, \au{Kwon, J.M.}, \au{Lee, J.P.}, \au{Lee, W.C.}, \au{Jeon, Y.M.}, \au{Kim, K.}, \au{Lee, J.H.}, \au{Shin, G.W.}, \au{Kim, J.}, \au{Lee, J.}, \au{Hahn, S.H.}, \au{Lee, J.W.}, \au{Kim, H.S.}, \au{Bak, J.G.}, \au{Lee, S.G.}, \au{Lee, Y.H.}, \au{Jeong, J.H.}, \au{Woo, M.H.}, \au{Kim, J.H.}, \au{Juhn, J.W.}, \au{Ko, J.S.}, \au{Sung, C.}, \au{Shin, H.W.}, \au{Park, J.M.}, \au{Kim, S.K.}, \au{Park, J.K.}, \au{Logan, N.C.}, \au{Yang, S.M.}, \au{Kolemen, E.}, \au{Hu, Q.M.}, \au{Shousha, R.}, \au{Barr, J.}, \au{Paz-Soldan, C.}, \au{Park, Y.S.}, \au{Sabbagh, S.A.}, \au{Ida, K.}, \au{Kim, S.}, \au{Loarte, A.}, \au{Gilson, E.}, \au{Eldon, D.}, \au{Nakano, T.}, \au{Tala, T.} \& \au{Team, {KSTAR}}}
  \yr{2024}  \at{Overview of the {KSTAR} experiments toward fusion reactor}.  \jt{Nuclear Fusion}  \bvol{64}~(11),  \pg{112010}.

\bibitem[Krupka(2024)]{Thesis}
{\sc \au{Krupka, Anna}} \yr{2024}  \at{Plasma speed optimization for improved tokamak plasma confinement}. PhD thesis, thèse de doctorat dirigée par Firpo, Marie Christine Physique Institut polytechnique de Paris 2024.

\bibitem[Krupka \& Firpo(2024)]{KRUPKA_scaling}
{\sc \au{Krupka, A.} \& \au{Firpo, M.-C.}} \yr{2024}  \at{Scaling laws of the plasma velocity in visco-resistive magnetohydrodynamic systems}.  \jt{Fundamental Plasma Physics}  \bvol{10},  \pg{100044}.

\bibitem[Li \& Zhu(2021)]{Li2021}
{\sc \au{Li, Haolong} \& \au{Zhu, Ping}} \yr{2021}  \at{Solving the {{Grad–Shafranov}} equation using spectral elements for tokamak equilibrium with toroidal rotation}.  \jt{Computer Physics Communications}  \bvol{260},  \pg{107264}.

\bibitem[Litaudon {\em et~al.\/}(2002)Litaudon, Crisanti, Alper, Artaud, Baranov, Barbato, Basiuk, Bécoulet, Bécoulet, Castaldo, Challis, Conway, Dux, Eriksson, Esposito, Fourment, Frigione, Garbet, Giroud, Hawkes, Hennequin, Huysmans, Imbeaux, Joffrin, Lomas, Lotte, Maget, Mantsinen, Mailloux, Mazon, Milani, Moreau, Parail, Pohn, Rimini, Sarazin, Tresset, Zastrow, Zerbini \& contributors to~the EFDA-JET~Workprogramme]{Litaudon_2002}
{\sc \au{Litaudon, X}, \au{Crisanti, F}, \au{Alper, B}, \au{Artaud, J~F}, \au{Baranov, Yu~F}, \au{Barbato, E}, \au{Basiuk, V}, \au{Bécoulet, A}, \au{Bécoulet, M}, \au{Castaldo, C}, \au{Challis, C~D}, \au{Conway, G~D}, \au{Dux, R}, \au{Eriksson, L~G}, \au{Esposito, B}, \au{Fourment, C}, \au{Frigione, D}, \au{Garbet, X}, \au{Giroud, C}, \au{Hawkes, N~C}, \au{Hennequin, P}, \au{Huysmans, G T~A}, \au{Imbeaux, F}, \au{Joffrin, E}, \au{Lomas, P~J}, \au{Lotte, Ph}, \au{Maget, P}, \au{Mantsinen, M}, \au{Mailloux, J}, \au{Mazon, D}, \au{Milani, F}, \au{Moreau, D}, \au{Parail, V}, \au{Pohn, E}, \au{Rimini, F~G}, \au{Sarazin, Y}, \au{Tresset, G}, \au{Zastrow, K~D}, \au{Zerbini, M} \& \au{contributors to~the EFDA-JET~Workprogramme}} \yr{2002}  \at{Towards fully non-inductive current drive operation in jet}.  \jt{Plasma Physics and Controlled Fusion}  \bvol{44}~(7),  \pg{1057}.

\bibitem[Montgomery(1993)]{Montgomery1993}
{\sc \au{Montgomery, D}} \yr{1993}  \at{Hartmann, {{Lundquist, and Reynolds}}: the role of dimensionless numbers in nonlinear magnetofluid behavior}.  \jt{Plasma Physics and Controlled Fusion}  \bvol{35}~(SB),  \pg{B105}.

\bibitem[Morales {\em et~al.\/}(2012)Morales, Bos, Schneider \& Montgomery]{Moraler2012}
{\sc \au{Morales, Jorge~A.}, \au{Bos, Wouter J.~T.}, \au{Schneider, Kai} \& \au{Montgomery, David~C.}} \yr{2012}  \at{Intrinsic rotation of toroidally confined magnetohydrodynamics}.  \jt{Phys. Rev. Lett.}  \bvol{109},  \pg{175002}.

\bibitem[Oueslati {\em et~al.\/}(2019)Oueslati, Bonnet, Minesi, Firpo \& Salhi]{oueslati2019numerical}
{\sc \au{Oueslati, H.}, \au{Bonnet, T.}, \au{Minesi, N.}, \au{Firpo, M.-C.} \& \au{Salhi, A.}} \yr{2019}  \at{Numerical derivation of steady flows in visco-resistive magnetohydrodynamics for {{JET}} and {{ITER-like}} geometries with no symmetry breaking}.  \jt{AIP Conference Proceedings}  \bvol{2179},  \pg{020009}.

\bibitem[Oueslati \& Firpo(2020)]{oueslati2020breaking}
{\sc \au{Oueslati, H.} \& \au{Firpo, M.-C.}} \yr{2020}  \at{Breaking up-down symmetry with magnetic perturbations in tokamak plasmas: {{Increase}} of axisymmetric steady-state velocities}.  \jt{Physics of Plasmas}  \bvol{27}~(10),  \pg{102501}.

\bibitem[Roverc'h {\em et~al.\/}(2021)Roverc'h, Oueslati \& Firpo]{roverchSteadystateFlowsViscoresistive2021}
{\sc \au{Roverc'h, E.}, \au{Oueslati, H.} \& \au{Firpo, M.-C.}} \yr{2021}  \at{Steady-state flows in a visco-resistive magnetohydrodynamic model of tokamak plasmas with inhomogeneous heating}.  \jt{Journal of Plasma Physics}  \bvol{87}~(2),  \pg{905870217}.

\bibitem[Sauter {\em et~al.\/}(2000)Sauter, Henderson, Hofmann, Goodman, Alberti, Angioni, Appert, Behn, Blanchard, Bosshard, Chavan, Coda, Duval, Fasel, Favre, Furno, Gorgerat, Hogge, Isoz, Joye, Lavanchy, Lister, Llobet, Magnin, Mandrin, Manini, Marl\'etaz, Marmillod, Martin, Mayor, Martynov, Mlynar, Moret, Nieswand, Nikkola, Paris, Perez, Pietrzyk, Pitts, Pochelon, Pochon, Refke, Reimerdes, Rommers, Scavino, Tonetti, Tran, Troyon \& Weisen]{Sauter_2000}
{\sc \au{Sauter, O.}, \au{Henderson, M.~A.}, \au{Hofmann, F.}, \au{Goodman, T.}, \au{Alberti, S.}, \au{Angioni, C.}, \au{Appert, K.}, \au{Behn, R.}, \au{Blanchard, P.}, \au{Bosshard, P.}, \au{Chavan, R.}, \au{Coda, S.}, \au{Duval, B.~P.}, \au{Fasel, D.}, \au{Favre, A.}, \au{Furno, I.}, \au{Gorgerat, P.}, \au{Hogge, J.-P.}, \au{Isoz, P.-F.}, \au{Joye, B.}, \au{Lavanchy, P.}, \au{Lister, J.~B.}, \au{Llobet, X.}, \au{Magnin, J.-C.}, \au{Mandrin, P.}, \au{Manini, A.}, \au{Marl\'etaz, B.}, \au{Marmillod, P.}, \au{Martin, Y.}, \au{Mayor, J.-M.}, \au{Martynov, A.~A.}, \au{Mlynar, J.}, \au{Moret, J.-M.}, \au{Nieswand, C.}, \au{Nikkola, P.}, \au{Paris, P.}, \au{Perez, A.}, \au{Pietrzyk, Z.~A.}, \au{Pitts, R.~A.}, \au{Pochelon, A.}, \au{Pochon, G.}, \au{Refke, A.}, \au{Reimerdes, H.}, \au{Rommers, J.}, \au{Scavino, E.}, \au{Tonetti, G.}, \au{Tran, M.~Q.}, \au{Troyon, F.} \& \au{Weisen, H.}} \yr{2000}  \at{Steady-state fully noninductive current driven by electron cyclotron waves in a magnetically confined plasma}.
  \jt{Phys. Rev. Lett.}  \bvol{84},  \pg{3322--3325}.

\bibitem[Tran {\em et~al.\/}(2022)Tran, Agostinetti, Aiello, Avramidis, Baiocchi, Barbisan, Bobkov, Briefi, Bruschi, Chavan, Chelis, Day, Delogu, Ell, Fanale, Fassina, Fantz, Faugel, Figini, Fiorucci, Friedl, Franke, Gantenbein, Garavaglia, Granucci, Hanke, Hogge, Hopf, Kostic, Illy, Ioannidis, Jelonnek, Jin, Latsas, Louche, Maquet, Maggiora, Messiaen, Milanesio, Mimo, Moro, Ochoukov, Ongena, Pagonakis, Peponis, Pimazzoni, Ragona, Rispoli, Ruess, Rzesnicki, Scherer, Spaeh, Starnella, Strauss, Thumm, Tierens, Tigelis, Tsironis, Usoltceva, {Van Eester}, Veronese, Vincenzi, Wagner, Wu, Zeus \& Zhang]{TRAN2022}
{\sc \au{Tran, M.Q.}, \au{Agostinetti, P.}, \au{Aiello, G.}, \au{Avramidis, K.}, \au{Baiocchi, B.}, \au{Barbisan, M.}, \au{Bobkov, V.}, \au{Briefi, S.}, \au{Bruschi, A.}, \au{Chavan, R.}, \au{Chelis, I.}, \au{Day, Ch.}, \au{Delogu, R.}, \au{Ell, B.}, \au{Fanale, F.}, \au{Fassina, A.}, \au{Fantz, U.}, \au{Faugel, H.}, \au{Figini, L.}, \au{Fiorucci, D.}, \au{Friedl, R.}, \au{Franke, Th.}, \au{Gantenbein, G.}, \au{Garavaglia, S.}, \au{Granucci, G.}, \au{Hanke, S.}, \au{Hogge, J.-P.}, \au{Hopf, C.}, \au{Kostic, A.}, \au{Illy, S.}, \au{Ioannidis, Z.}, \au{Jelonnek, J.}, \au{Jin, J.}, \au{Latsas, G.}, \au{Louche, F.}, \au{Maquet, V.}, \au{Maggiora, R.}, \au{Messiaen, A.}, \au{Milanesio, D.}, \au{Mimo, A.}, \au{Moro, A.}, \au{Ochoukov, R.}, \au{Ongena, J.}, \au{Pagonakis, I.G.}, \au{Peponis, D.}, \au{Pimazzoni, A.}, \au{Ragona, R.}, \au{Rispoli, N.}, \au{Ruess, T.}, \au{Rzesnicki, T.}, \au{Scherer, T.}, \au{Spaeh, P.}, \au{Starnella, G.}, \au{Strauss, D.}, \au{Thumm, M.}, \au{Tierens, W.}, \au{Tigelis, I.},
  \au{Tsironis, C.}, \au{Usoltceva, M.}, \au{{Van Eester}, D.}, \au{Veronese, F.}, \au{Vincenzi, P.}, \au{Wagner, F.}, \au{Wu, C.}, \au{Zeus, F.} \& \au{Zhang, W.}} \yr{2022}  \at{Status and future development of heating and current drive for the {EU DEMO}}.  \jt{Fusion Engineering and Design}  \bvol{180},  \pg{113159}.

\bibitem[Wan {\em et~al.\/}(2017)Wan, Li, Liu, Wang, Chan, Chen, Duan, Fu, Gao, Feng, Liu, Song, Weng, Wan, Wan, Wang, Wu, Ye, Yang, Zheng, Zhuang, Li \& team]{Wan_2017}
{\sc \au{Wan, Yuanxi}, \au{Li, Jiangang}, \au{Liu, Yong}, \au{Wang, Xiaolin}, \au{Chan, Vincent}, \au{Chen, Changan}, \au{Duan, Xuru}, \au{Fu, Peng}, \au{Gao, Xiang}, \au{Feng, Kaiming}, \au{Liu, Songlin}, \au{Song, Yuntao}, \au{Weng, Peide}, \au{Wan, Baonian}, \au{Wan, Farong}, \au{Wang, Heyi}, \au{Wu, Songtao}, \au{Ye, Minyou}, \au{Yang, Qingwei}, \au{Zheng, Guoyao}, \au{Zhuang, Ge}, \au{Li, Qiang} \& \au{team, {CFETR}}} \yr{2017}  \at{Overview of the present progress and activities on the {CFETR}}.  \jt{Nuclear Fusion}  \bvol{57}~(10),  \pg{102009}.

\bibitem[Xie {\em et~al.\/}(2023)Xie, Hu, Liu, Xu, Wei, zhao, Jiang, Liang, Pan, Li, Yi \& Xie]{NBI_EAST_2023}
{\sc \au{Xie, Yahong}, \au{Hu, Chundong}, \au{Liu, Sheng}, \au{Xu, Yongjian}, \au{Wei, Jianglong}, \au{zhao, Yuanzhe}, \au{Jiang, Caichao}, \au{Liang, Lizhen}, \au{Pan, Junjun}, \au{Li, Jun}, \au{Yi, Wei} \& \au{Xie, Yuanlai}} \yr{2023}  \at{Long pulse operation of neutral beam injector on {EAST} tokamak}.  \jt{Fusion Engineering and Design}  \bvol{193},  \pg{113744}.

\end{thebibliography}
\end{document}